\documentclass[%
reprint,
superscriptaddress,
showpacs,preprintnumbers,
amsmath,amssymb,
prb
]{revtex4-1}

\usepackage[pdftex]{graphicx}
\usepackage[pdftex]{color}

\usepackage{dcolumn}
\usepackage{bm}

\usepackage{color} 

\begin{document}

\preprint{APS/123-QED}

\title{Laser induced surface magnetization in Floquet-Weyl semimetals}

\author{Runnan Zhang}
\affiliation{Graduate School of Science and Technology, University of Tsukuba,  Tsukuba, Ibaraki 305-8573, Japan}
\author{Ken-ichi Hino}
\email{hino@ims.tsukuba.ac.jp}
\affiliation{Institute of Pure and Applied Sciences, University of Tsukuba, Tsukuba, Ibaraki 305-8573, Japan}
\affiliation{Center for Computational Sciences, University of Tsukuba, Tsukuba, Ibaraki 305-8577, Japan}
\author{Nobuya Maeshima}
\affiliation{Center for Computational Sciences, University of Tsukuba, Tsukuba, Ibaraki  305-8577, Japan}
\author{Haruki Yogemura}
\affiliation{Graduate School of Science and Technology, University of Tsukuba, Tsukuba, Ibaraki  305-8573, Japan}
\author{Takeru Karikomi}
\affiliation{Graduate School of Science and Technology, University of Tsukuba, Tsukuba,  Ibaraki  305-8573, Japan}

\date{\today}

\begin{abstract}

We investigate optically induced magnetization in 
Floquet-Weyl semimetals generated by irradiation of a circularly-polarized continuous-wave laser from the group II-V narrow gap semiconductor Zn$_3$As$_2$ in a theoretical manner.
Here, this trivial and nonmagnetic crystal is driven by the laser with a nearly resonant frequency with a band gap to generate two types of Floquet-Weyl semimetal phases composed of different spin states.
These two phases host nontrivial two-dimensional surface states pinned to the respective pairs of the Weyl points.
By numerically evaluating the laser-induced transient carrier-dynamics, it is found that both spins are distributed in an uneven manner on the corresponding surface states due to significantly different excitation probabilities caused by the circularly-polarized laser with the nearly resonant frequency.
It is likely that such spin-polarized surface states produce surface magnetization, and furthermore the inverse Faraday effect also contributes almost as much as the spin magnetization. 
To be more specific, excited carries with high density of the order of $10^{21}\: {\rm cm}^{-3}$ are generated by the laser with electric field strength of a few MV/cm to result in the surface magnetization that becomes asymptotically constant with respect to time, around 1 mT.
The magnitude and the direction of it depend sharply on both of the intensity and frequency of the driving laser, which would be detected by virtue of the magneto-optic Kerr effect.

\end{abstract}

\maketitle


\section{Introduction}
\label{sec1}

Weyl semimetals (WSMs) that belong to a class of topological semimetals (SMs) \cite{Murakami2007a,Murakami2007b,Wang2012,Wang2013,Young2012,Young2015,Young2015,Yang2014,Yi2014,Xu2015,Kargarian2016,Park2017,Doh2017,Yan2017a,Armitage2018,Yang2018} have been extensively studied from the perspective of the nontrivial topology leading to many intriguing features of unusual electronic, magnetic, and optical properties.
Low-energy excitation of WSMs is described by the Weyl equation characteristic of a linearly dispersive gapless band and well-defined chirality, where the Berry curvature shows a monopole at the resulting crossing point termed the Weyl node. 
The WSMs result in fundamental properties such as Fermi arc surface states, axion electrodynamics, and chiral anomaly (the Adler-Bell-Jackiw anomaly) proposed in particle physics.\cite{Weyl1929,Majorana1937,Elliott2015,Adler1969,Bell1969,Nielsen1983} 
Further, these exhibit a great number of novel transport properties\cite{Yan2017a,Armitage2018} such as ultra-high mobility,\cite{Shekhar2015} titanic magnetic resistance,\cite{Shekhar2015} anomalous Hall conductivity,\cite{Wang2013,Gynther2011,Son2013,Arnold2016,Huang2015,Ali2014} and large negative magnetic resistance observed in magnetotransport attributed to the chiral anomaly. \cite{Son2013,Burkov2014,Burkov2015,Xiong2015,Zhang2016a}
The WSMs are created by breaking of either time reversal symmetry or spatial-inversion symmetry from Dirac SMs and nodal-line SMs that are different types of topological SMs.\cite{Yang2018,Armitage2018} 
In the WSMs with the time reversal symmetry breaking, there are a pair of Weyl nodes with opposite chirality, while in those with the spatial-inversion symmetry breaking, the number of Weyl nodes is a multiple of four. \cite{Murakami2007b,Armitage2018}

The introduction of an interaction of electron with a monochromatic continuous-wave laser with circular polarization also breaks the time reversal symmetry of the Dirac and nodal-line SMs to lead to a WSM state in nonequilibrium.\cite{Wang2014,Taguchi2016a,Taguchi2016b,Hubener2017,Zou2016,Zhang2016b,Yan2016,Yan2017b}
Due to the temporal periodicity of the Hamiltonian of concern, the Floquet theorem ensures the existence of a set of quasienergy bands (Floquet bands) associated with the WSM states, termed as Floquet WSM (FWSM) states.
The physical properties of the FWSMs can be controlled by means of built-in laser parameters such as intensity, frequency, and polarization, giving rise to band modulation and topological phase transitions.
The creation and optical control of Floquet states, termed as the Floquet engineering, provide a new avenue for the quantum control of topological materials.\cite{Goldman2014,Plekhanov2016,Oka2019,Rudner2020}
In passing, the spatial inversion symmetry is also broken by the application of the monochromatic continuous-wave laser.
However, the time-glide spatial inversion symmetry holds correct instead to realize the same invariance in the Hamiltonian as the spatial inversion symmetry.\cite{Zhang2021,Morimoto2017}

In recent years, there has been a growing interest in the FWSMs for exploring new topological properties that cannot be realized in WSMs in equilibrium. \cite{Chen2018,Kumar2019,Zhu2020,Salerno2020,Gao2020,Kawaguchi2020,Tokman2020,Liang2021,Ma2015,Juan2017,Ghorashi2018,Umer2021a,Umer2021b,Zhang2022,Yoshikawa2022,Guo2023}
It is found that the drive of a three-dimensional Luttinger SM by an elliptically polarized light leads to the coexistence of WSM phases with double and single Weyl points, which can be tuned to be type-I or type-II,\cite{Ghorashi2018} and the periodical driving of hybrid multi-WSM phases modulates the number of various isolated band touching points on-demand by tuning system parameters.\cite{Umer2021a}
Further, chirality switching of Weyl nodes is observed in a ferromagnetic Weyl semimetal excited by circularly polarized mid-infrared light pulses, which accompanies the reversal of magnetization.\cite{Yoshikawa2022}

It has been reported very recently that two different spin-polarized FWSM phases are created at one time by irradiating on a trivial narrow gap semiconductor, Zn$_3$As$_2$, an intense circularly-polarized laser with frequency that is almost on-resonant with the bandgap.\cite{Zhang2022} 
This is due to the fact that an electric-dipole transition induced by this laser between valence and conduction bands with one spin state is dominant over that between bands with the other spin state.
To be more specific, given the driving laser with a left-hand circular polarization (rotating in counterclockwise sense with time), the dipole coupling between the down-spin bands is much greater than that between the up-spin bands.
Thus, two types of Floquet bands with distinct energy structures are generated by the hybridization of valence and conduction bands with each spin state.
In particular, the resulting up-spin Floquet band forms just a considerably smaller band gap than that of the down-spin Floquet band, similar to a nodal line SM with a band crossing in the shape of a closed ring.
Furthermore, the associated surface states supported by these two Floquet bands also differ a lot from each other in energy structure.
This finding is much contrasted with properties of the FWSMs induced by a circularly polarized laser with an off-resonant frequency,\cite{Wang2014,Hubener2017,Hansen2017} in which a high-frequency approximation based on the Floquet-Magnus expansion \cite{Casas2001,Mananga2011,Eckardt2015,Mananga2016,Haga2019} is warranted.

This paper is aimed at examining the possibility of the manifestation of transient magnetization attributed to the above-stated difference of the dipole couplings in the FWSM created by the on-resonant drive.
It is stressed that real carrier excitations caused by such transient quantum dynamics play crucial roles of the magnetization, differing from the off-resonant drive just causing virtual excitations.
Due to the application of the intense laser with a peak electric field of the order of 1~MV/cm, down-spin electrons are considered to be dominantly excited than up-spin electrons, resulting in the imbalance of spin distributions between the respective Floquet bands.
It is likely that conspicuous surface magnetization arises from such imbalance of spin distributions, namely, the surface spin polarization; where the associated bulk magnetization is also induced, however actually neglected compared with the surface one, as shown later (Sect.\ref{sec3}).
Furthermore, it is considered that the inverse Faraday effect (IFE) contributes to the surface magnetization to some extent in addition with the above spin magnetization.
The IFE is given rise to by the laser with circular polarization, and the orbital magnetization is generated, which remains constant in time.\cite{Gao2020,Kawaguchi2020,Tokman2020,Liang2021,Pershan1966,Kimel2005,Vahaplar2009,Ostler2012,Hertel2006,Zhang2009,Woodford2009,Kurkin2008,Taguchi2011,Popova2011,Popova2012,Gridnev2013,Qaiumzadeh2013,Battiato2014}
Differing from the conventional IFE arising from the second-order optical process with respect to the electric field of the laser, it is expected that the higher order terms come into play for the orbital magnetization concerned here due to the application of the intense laser.
In passing, it is remarked that the IFE on WSMs, differing from the present FWSMs, has been studied recently to exhibit frequency-independent magnetization mechanisms. \cite{Gao2020,Tokman2020}

The remainder of this paper is organized as follows. 
Section~\ref{sec2} describes the theoretical framework, Sec.~\ref{sec3} presents the results and discussion, and Sec.~\ref{sec4} presents the conclusion.
Further, three appendices are included. 
Hereafter, the atomic units (a.u.) are used throughout, unless otherwise stated.

\section{Theory}
\label{sec2}

In this section, one begins with the construction of the Floquet Hamiltonian in Sec.~\ref{sec2A}, followed by the description of the density matrices for the laser induced dynamics of concern in Sec.~\ref{sec2B} and of the laser induced magnetization in Sec.~\ref{sec2C}.

\subsection{Floquet Hamiltonian}
\label{sec2A}

The crystal adopted here is the group II-V narrow gap semiconductor, Zn$_3$As$_2$, with the C$_4$-rotational symmetry along the $\Gamma-$Z axis in the the Brillouin zone (BZ).
This structure is similar to that of Cd$_3$As$_2$, though the band of the latter is inverted to result in a Dirac SM.\cite{LinChung1969,Sieranski1994,Neupane2014,Liu2014,Crassee2018}
Here, one employs the structure of $\alpha^\prime$Zn$_3$As$_2$ with a tetragonal structure P4$_2/$nbc \cite{Okamoto1992} among many equilibrium phases of Zn$_3$As$_2$ depending on pressure and temperature.

An effective electronic Hamiltonian of this crystal is constructed based on the Kane model\cite{Wang2013,Zhang2022} in view of the low-energy electronic properties mostly determined by the conduction band composed of Zn 4$s$-orbitals and the valence band composed of As 4$p$-orbitals.
Thus, the following four states are considered as conduction $s$-states
$|\Gamma_6, J_z=\pm 1/2\rangle$ and heavy-hole $p$-states $|\Gamma_7, J_z=\pm 3/2\rangle$, where $J_z$ represents the $z$ component of total angular momentum quantum number at the $\Gamma$-point with a 3D Bloch momentum $\boldsymbol{k}=(k_x, k_y,k_z)$ being equal to zero.
Here, light-hole states and split-off states are disregarded because of relatively large energy separation from these four states at the $\Gamma$-point.
The effective Hamiltonian is given by the $4 \times 4$-matrix:
\cite{Kane1957,Luttinger1956}
\begin{equation}
\mathcal{H}(\boldsymbol{k})
=\sum_{j=3}^5 D_j(\boldsymbol{k})\gamma_j,
\label{calH}
\end{equation}
where $\gamma_j$'s represent the four-dimensional Dirac matrices for the Clifford algebra ensuring the anti-commutation relation,
\(
\{\gamma_j, \gamma_{j^\prime} \}=2\delta_{jj^\prime}
\),
defined by
\(
\gamma_1=\tau_x\otimes\sigma_x,\:
\gamma_2=\tau_x\otimes\sigma_y,\:
\gamma_3=\tau_x\otimes\sigma_z,\:
\gamma_4=\tau_z\otimes I_2
\), and 
\(
\gamma_5=\tau_y\otimes I_2
\);
$I_2$ represents the $2\times 2$ unit matrix, and $\tau_l$ and $\sigma_l$ with $l=x,\:y,\:z$ represent the Pauli matrices for orbital and spin degrees of freedom, respectively.
According to the above definition of $\gamma_j$'s, it is understood that
the states of
$|\Gamma_6, J_z=1/2\rangle,\: |\Gamma_7, J_z= 3/2\rangle,\:
|\Gamma_6, J_z=-1/2\rangle$,
and
$|\Gamma_7, J_z=-3/2\rangle
$
are labeled as $1, 2, 3$, and 4, respectively,
for the matrix elements of
$\mathcal{H}(\boldsymbol{k})$, namely, $\{\mathcal{H}_{mn}(\boldsymbol{k})\}$ 
with $m,n=1\sim4$.
Moreover, $D_j(\boldsymbol{k})$'s are given by
\begin{eqnarray}
\left\{
\begin{array}{l}
D_3(\boldsymbol{k})=t_{sp}\sin{(k_xd_x)}\\
D_4(\boldsymbol{k})=\Delta_g
-2\sum_{l=x,y,z}t^{l}\{1-\cos{(k_ld_l)}\}\\
D_5(\boldsymbol{k})=t_{sp}\sin{(k_yd_y)}
\end{array}
\right..
\label{dk}
\end{eqnarray}
Here, $t^l$ represents a hopping matrix between identical bands in the $l$-direction
with $l=x,y,z$ where $t^l<0$ and $t^x=t^y$, and $t_{sp}$ represents a hopping matrix between different bands due to a spin-orbit coupling.
Further, $d_l$ represents a lattice constant in the $l$-direction, and the bandgap at the $\Gamma$-point $E_g$ is given by the difference of the energy of conduction band at $\Gamma_6$ from that of valence band at $\Gamma_7$ with $\Delta_g=E_g/2$.

An interaction of electron with light is introduced into $\mathcal{H}(\boldsymbol{k})$
by replacing $\boldsymbol{k}$ by $\boldsymbol{K}(t)=\boldsymbol{k}+\boldsymbol{A}(t)$, followed by adding to $\mathcal{H}(\boldsymbol{K}(t))$ an interband electric-dipole interaction represented by
\(
\mathcal{H}^\prime(t).
\)
Here, $\boldsymbol{A}(t)$ represents a vector potential of the laser at time $t$, given by
\begin{equation}
\boldsymbol{A}(t)=\left(
-{F_x\over \omega} \sin\omega t,{F_y\over \omega} \cos\omega t,0
\right)
\label{A}
\end{equation}
with $\omega$ as frequency, and $F_x$ and $F_y$ as constants, and in view of $\boldsymbol{F}(t)=-\dot{\boldsymbol{A}}(t)$, the associated electric field becomes
\begin{equation}
\boldsymbol{F}(t)=\left(
F_x\cos\omega t,F_y\sin\omega t,0
\right).
\label{F}
\end{equation}
The laser is linearly polarized in the $x$-direction when $F_x\not=0$ and $F_y=0$, while left-hand circularly polarized in the $x-y$ plane when $F_x$ is set equal to $F_y$, namely, $F_c\equiv F_x=F_y$.
The time-dependent effective Hamiltonian of the driven semiconductor is thus read as
\begin{equation}
H_F(\boldsymbol{k},t)=\mathcal{H}(\boldsymbol{K}(t))
+\mathcal{H}^\prime(t),
\label{HF}
\end{equation}
where
\begin{equation}
\mathcal{H}(\boldsymbol{K}(t)=\sum_{j=3}^5 D_j(\boldsymbol{K}(t))\gamma_j.
\end{equation}
The interband electric-dipole interaction is provided as\cite{Zhang2022} 
\begin{equation}
\mathcal{H}^\prime(t)
=(\Omega_y\sin{\omega t})\tau_x\otimes I_2
+(\Omega_x\cos{\omega t}) \tau_y\otimes \sigma_z,
\label{Hprime}
\end{equation}
where
$\Omega_x=F_xP/\sqrt{2}$ and $\Omega_y=F_yP/\sqrt{2}$.
Here, $P$ is a dipole matrix element given by
$P=\langle S|x|X\rangle =\langle S|y|Y\rangle$,
where
$x$ and $y$ represent the $x$ and $y$ components of electron position $\boldsymbol{r}$,
respectively,
and the states of
$|\Gamma_6, J_z=\pm 1/2\rangle$ and $|\Gamma_7, J_z=\pm 3/2\rangle$ are
represented by
$|\Gamma_6, J_z=\pm 1/2\rangle=i|S\rangle$
and
$|\Gamma_7, J_z=\pm 3/2\rangle=\pm(1/\sqrt{2})|X\pm i Y\rangle$,
respectively,
in terms of $s, p_x$, and $p_y$ states denoted as
$|S\rangle, |X\rangle$, and $|Y\rangle$, respectively.

Obviously, this Hamiltonian ensures the temporal periodicity, $H_F(\boldsymbol{k},t+T)=H_F(\boldsymbol{k},t)$
with $T=2\pi/\omega$, and the system of concern follows the Floquet theorem.\cite{Shirley1965}
Thus, the present time-dependent problem ends up with the following Floquet eigenvalue problem as:
\begin{equation}
L_F(\boldsymbol{k},t)\phi_{\boldsymbol{k}\alpha}(t)=E_\alpha(\boldsymbol{k})\phi_{\boldsymbol{k}\alpha}(t),
\label{Floquet}
\end{equation}
where $L_F(\boldsymbol{k},t)$ is the Floquet Hamiltonian defined as
\begin{equation}
L_F(\boldsymbol{k},t)=H_F(\boldsymbol{k},t)-iI{\partial\over \partial t}
\label{LF}
\end{equation}
with $I$ the $4 \times 4$ unit matrix, $E_\alpha(\boldsymbol{k})$ represents an eigenvalue showing the $\alpha$th Floquet energy, and $\phi_{\boldsymbol{k}\alpha}(t)$ represents the associated eigenvector ensuring the temporal periodicity, $\phi_{\boldsymbol{k}\alpha}(t+T)=\phi_{\boldsymbol{k}\alpha}(t)$.
Hereafter, it is understood that the photon sideband $\alpha$ is represented as $\alpha_0(N_\alpha)$, where $\alpha_0(\equiv\alpha_0(0))$ and $N_\alpha$ stand for a parent band and the number of the sideband, respectively.
Further, $\alpha(N)\equiv\alpha_0(N_\alpha+N)$ and $E_{\alpha(N)}(\boldsymbol{k})\equiv E_\alpha(\boldsymbol{k})+N\omega=E_{\alpha_0}(\boldsymbol{k})+(N_\alpha+N)\:\omega$ with $N$ being an integer.

For the purpose of the later convenience, the wavefunction $\phi_{\boldsymbol{k}\alpha}(t)$ is represented in view of the above periodicity as
\begin{equation}
\phi_{\boldsymbol{k}\alpha}(t)=\sum_\mu \chi_\mu \varphi_{\mu,\boldsymbol{k}\alpha}(t),
\label{phi0}
\end{equation}
where $\chi_{\mu}$ represents the $\mu$th state corresponding to $|\Gamma_6, J_z=\pm 1/2\rangle$ and $|\Gamma_7, J_z=\pm 3/2\rangle$, and the periodic function $\varphi_{\mu,\boldsymbol{k}\alpha}(t)$ is 
given in terms of the Fourier expansion as
\begin{equation}
\varphi_{\mu,\boldsymbol{k}\alpha}(t)=\sum_ne^{in\omega t}C_{\mu n,\boldsymbol{k}\alpha}.
\label{varphi2}
\end{equation}
Thus, Eq.~(\ref{Floquet}) is recast into
\begin{equation}
\sum_{\nu}
L_{F\mu\nu}(\boldsymbol{k},t)\varphi_{\nu,\boldsymbol{k}\alpha}(t)=E_\alpha(\boldsymbol{k})\varphi_{\mu,\boldsymbol{k}\alpha}(t),
\label{varphi1}
\end{equation}
where
\begin{equation}
L_{F\mu\nu}(\boldsymbol{k},t)= H_{F\mu\nu}(\boldsymbol{k},t)-\delta_{\mu\nu}i{d\over
dt},
\label{L2}
\end{equation}
and
\begin{equation}
\sum_{\nu n^\prime}L_{F\:\mu n,\nu n^\prime}(\boldsymbol{k})C_{\nu n^\prime,\boldsymbol{k}\alpha}
=E_\alpha(\boldsymbol{k}) C_{\mu n,\boldsymbol{k}\alpha},
\label{EqC}
\end{equation}
where
\begin{equation}
L_{F\:\mu n,\nu n^\prime}(\boldsymbol{k})
={1\over T}\int^T_0 dt\:e^{-in\omega t}
L_{F\mu\nu}(\boldsymbol{k},t) \: e^{in^\prime\omega t}.
\end{equation}

\subsection{Density matrices}
\label{sec2B}

In the actual physical system, the intense monochromatic laser with circular polarization is exerted on the crystal at a certain time, for example, at $t=0$, leading to high-density carrier excitation followed by dephasing and relaxation.
The detail of this transient dynamics with concomitant physical phenomena can be examined by tracking the alteration of the associated density matrices in time.
The one-electron Hamiltonian representing the dynamics is given by
\begin{equation}
H(\boldsymbol{k},t)=\mathcal{H}(\boldsymbol{k})+V(\boldsymbol{k},t),
\label{H}
\end{equation}
where
\begin{equation}
V(\boldsymbol{k},t)\equiv
[H_F(\boldsymbol{k},t)-\mathcal{H}(\boldsymbol{k})]\theta(t)
\end{equation}
under the tacit assumption that temporal width of the applied laser pulse is sufficiently long.
Obviously, $H(\boldsymbol{k},t+T)\not=H(\boldsymbol{k},t)$, since $H(\boldsymbol{k},t)=\mathcal{H}(\boldsymbol{k})$ for $t<0$ and $H(\boldsymbol{k},t)=H_F(\boldsymbol{k},t)$ for $t>0$.
Employing Eq.~(\ref{H}), the total Hamiltonian represented by $\hat{H}(t)$ is read as
\begin{equation}
\hat{H}(t)=\sum_{\boldsymbol{k}\mu\nu} H_{\mu\nu}(\boldsymbol{k},t)\:
\hat{a}^\dagger_{\boldsymbol{k}\mu}(t)\hat{a}_{\boldsymbol{k}\nu}(t),
\label{Hhat}
\end{equation}
where $\hat{a}^\dagger_{\boldsymbol{k}\mu}(t)$ and $\hat{a}_{\boldsymbol{k}\mu}(t)$ represent
creation and annihilation operators of state $\mu$ at $\boldsymbol{k}$ in the
Heisenberg representation, respectively, and
electron-electron interactions are neglected for the sake of
simplicity.
Furthermore, for a given $\boldsymbol{k}$, the field operator of the system is provided by
\begin{equation}
\hat{\psi}_{\boldsymbol{k}}(t)=\sum_\mu \chi_{\mu}\hat{a}_{\boldsymbol{k}\mu}(t).
\label{Psihat}
\end{equation}

The time-evolution of operator $\hat{a}_{\boldsymbol{k}\lambda}(t)$ follows the Heisenberg equation of
\begin{eqnarray}
i{d \hat{a}_{\boldsymbol{k}\lambda}(t) \over dt}
&=&
[\hat{a}_{\boldsymbol{k}\lambda}(t),
\hat{H}(t)]
\nonumber\\
&=&
\sum_{\nu} H_{\lambda\nu}(\boldsymbol{k},t) \hat{a}_{\boldsymbol{k}\nu}(t),
\label{Eqahat}
\end{eqnarray}
where the equal-time anti-commutation relation $\{\hat{a}_{\boldsymbol{k}\mu}(t),\hat{a}^\dagger_{\boldsymbol{k}^\prime\nu}(t)\}=\delta_{\mu\nu}\delta_{\boldsymbol{k}\boldsymbol{k}^\prime}$ is employed.
Hence, it is shown that a reduced density matrix, defined by
\begin{equation}
\rho_{\boldsymbol{k}\mu\nu}(t)=\langle \hat{a}^\dagger_{\boldsymbol{k}\mu}(t)\hat{a}_{\boldsymbol{k}\nu}(t)
\rangle_g,
\label{rhoahat}
\end{equation}
ensures the equation of motion given by
\begin{eqnarray}
i{d\rho_{\boldsymbol{k}\mu\nu}(t)\over dt}
&=&\Big\langle
[\hat{a}^\dagger_{\boldsymbol{k}\mu}(t)\hat{a}_{\boldsymbol{k}\nu}(t),\hat{H}(t)]
\Big\rangle_g
\nonumber\\
&=&
\sum_\lambda
\Big[
-H^*_{\mu\lambda}(\boldsymbol{k},t) \rho_{\boldsymbol{k}\lambda\nu}(t)
\nonumber\\
&&
+H^*_{\lambda\nu}(\boldsymbol{k},t) \rho_{\boldsymbol{k}\mu\lambda}(t)
\Big].
\label{OBeqs}
\end{eqnarray}
Hereafter, the reduced density matrix Eq.~(\ref{rhoahat}) is simply termed as a density matrix.
$\langle \hat{X} \rangle_g$ represents a ground state expectation value of operator $\hat{X}$.
In addition, the contributions of phenomenological damping have to be incorporated with Eq.~(\ref{OBeqs}), as is taken account of later.

Now, the operator $\hat{b}^\dagger_{\boldsymbol{k}\alpha}(t)$ is defined as
\begin{equation}
\hat{b}_{\boldsymbol{k}\alpha}^\dagger(t)=\sum_\mu\hat{a}_{\boldsymbol{k}\mu}^\dagger(t)\varphi_{\mu,\boldsymbol{k}\alpha}(t)\theta(t),
\label{bahat}
\end{equation}
which brings the operator $\hat{a}^\dagger_{\boldsymbol{k}\mu}(t)$ to
\begin{equation}
\hat{a}^\dagger_{\boldsymbol{k}\mu}(t)=
\sum_\alpha\hat{b}_{\boldsymbol{k}\alpha}^\dagger(t)C^*_{\mu0,\boldsymbol{k}\alpha},
\label{abhat}
\end{equation}
as is shown in Eq.~(\ref{Arhoa2rhob}).
Hence, the field operator of Eq.~(\ref{Psihat}) is rewritten as
\begin{eqnarray}
\hat{\psi}_{\boldsymbol{k}}(t)
&=&
\sum_{\mu\alpha}\chi_\mu
C_{\mu 0,\boldsymbol{k}\alpha}\hat{b}_{\boldsymbol{k}\alpha}(t)
\nonumber\\
&=&
\sum_{\alpha_0} \phi_{\boldsymbol{k}\alpha_0}(t)\hat{b}_{\boldsymbol{k}\alpha_0}(t),
\end{eqnarray}
where Eqs.~(\ref{phi0}), (\ref{AC1}), and (\ref{Abahat}) are employed.
The equal-time anti-commutation between $\hat{b}_{\boldsymbol{k}\alpha}(t)$ and
$\hat{b}^\dagger_{\boldsymbol{k}^\prime\alpha^\prime}(t)$
is evaluated in view of Eq.~(\ref{varphi2}) as
\begin{eqnarray}
&&\{\hat{b}_{\boldsymbol{k}\alpha}(t),\hat{b}^\dagger_{\boldsymbol{k}^\prime\alpha^\prime}(t)\}
\nonumber\\
&&=
\delta_{\boldsymbol{k}\boldsymbol{k}^\prime}\sum_{\mu}\varphi_{\mu,\boldsymbol{k}\alpha}\varphi^*_{\mu,\boldsymbol{k}\alpha^\prime}(t)
\nonumber\\
&&=
\delta_{\boldsymbol{k}\boldsymbol{k}^\prime}\sum_{\mu n \Delta n}e^{i\Delta n\omega t}C_{\mu
n,\boldsymbol{k}\alpha}C^*_{\mu n,\boldsymbol{k}\alpha^\prime(\Delta n)}
\nonumber\\
&&=
\delta_{\boldsymbol{k}\boldsymbol{k}^\prime}\sum_{\Delta n}e^{i\Delta n \omega t}\delta_{\alpha_0\alpha^\prime_0}
\delta_{N_\alpha,(N_{\alpha^\prime}+\Delta n)}
\nonumber\\
&&=
\delta_{\boldsymbol{k}\boldsymbol{k}^\prime}e^{i(N_\alpha-N_{\alpha^\prime})\omega
t}\delta_{\alpha_0\alpha^\prime_0}.
\end{eqnarray}
Thus, one obtains the anti-commutation relation that
\begin{equation}
\{\hat{b}_{\boldsymbol{k}\alpha_0}(t),\hat{b}^\dagger_{\boldsymbol{k}^\prime\alpha^\prime_0}(t)\}
=\delta_{\boldsymbol{k}\boldsymbol{k}^\prime}\delta_{\alpha_0\alpha^\prime_0},
\label{commbb}
\end{equation}
which implies that
$\hat{b}_{\boldsymbol{k}\alpha_0}(t)$ and $\hat{b}^\dagger_{\boldsymbol{k}^\prime\alpha^\prime_0}(t)$
play the role of fermion operators.

Next, the equation of motion of $\hat{b}_{\boldsymbol{k}\alpha_0}(t)$ is derived.
In view of Eq.~(\ref{Eqahat}) with Eq.~(\ref{bahat}) and (\ref{abhat}),
\begin{eqnarray}
&&i{d\hat{b}_{\boldsymbol{k}\alpha_0}(t)\over dt}
\nonumber\\
&=&
\sum_\nu
\Bigg(
i{d\hat{a}_{\boldsymbol{k}\nu}(t)\over dt}\varphi^*_{\nu,\boldsymbol{k}\alpha_0}(t)
+
\hat{a}_{\boldsymbol{k}\nu}(t)i{d\varphi^*_{\nu,\boldsymbol{k}\alpha_0}(t)\over dt}
\Bigg)
\nonumber\\
&&
+i\hat{b}_{\boldsymbol{k}\alpha_0}(0)\delta(t)
\nonumber
\\
&=&
[\hat{b}_{\boldsymbol{k}\alpha_0}(t),\hat{H}(t)]
+
\sum_\nu
\hat{a}_{\boldsymbol{k}\nu}(t)i{d\varphi^*_{\nu,\boldsymbol{k}\alpha_0}(t)\over dt}
\nonumber\\
&&
+i\hat{b}_{\boldsymbol{k}\alpha_0}(0)\delta(t).
\label{Eqbhat0}
\end{eqnarray}
On the other hand, defining another Hamiltonian as
\begin{equation}
\hat{L}(t)
=\sum_{\boldsymbol{k}\alpha_0}E_{\alpha_0}(\boldsymbol{k})\hat{b}^\dagger_{\boldsymbol{k}\alpha_0}(t)\hat{b}_{\boldsymbol{k}\alpha_0}(t),
\label{Lhat}
\end{equation}
and employing Eq.~(\ref{bahat}),
this is cast into
\begin{eqnarray}
\hat{L}(t)
&=&
\sum_{\boldsymbol{k}\alpha_0 n \mu\nu}
E_{\alpha_0}(\boldsymbol{k})\varphi_{\mu,\boldsymbol{k}\alpha_0}(t)e^{-in\omega t}C^*_{\nu n,\boldsymbol{k}\alpha_0}
\hat{a}^\dagger_{\boldsymbol{k}\mu}(t)\hat{a}_{\boldsymbol{k}\nu}(t)
\nonumber\\
&=&
\sum_{\boldsymbol{k}\alpha_0 n \mu\nu}
(E_{\alpha_0(-n)}(\boldsymbol{k})+n\omega)\varphi_{\mu,\boldsymbol{k}\alpha_0(-n)}(t)
\nonumber\\
&&\times
C^*_{\nu 0,\boldsymbol{k}\alpha_0(-n)}
\hat{a}^\dagger_{\boldsymbol{k}\mu}(t)\hat{a}_{\boldsymbol{k}\nu}(t)
\nonumber\\
&=&
\sum_{\boldsymbol{k}\mu\nu}H_{\mu\nu}(\boldsymbol{k},t)\hat{a}^\dagger_{\boldsymbol{k}\mu}(t)\hat{a}_{\boldsymbol{k}\nu}(t)
\nonumber\\
&&
+
\sum_{\boldsymbol{k}\alpha_0 n \nu}
n\omega C^*_{\nu 0,\boldsymbol{k}\alpha_0(-n)}
e^{-in\omega t}
\hat{b}^\dagger_{\boldsymbol{k}\alpha_0}(t)
\hat{a}_{\boldsymbol{k}\nu}(t)
\nonumber\\
&=&
\hat{H}(t)
+
\sum_{\boldsymbol{k}\alpha_0 \nu}
\hat{b}^\dagger_{\boldsymbol{k}\alpha_0}(t)
\hat{a}_{\boldsymbol{k}\nu}(t)
i{d \varphi^*_{\nu,\boldsymbol{k}\alpha_0}(t)\over d t},
\end{eqnarray}
where
Eqs.~(\ref{varphi2}), (\ref{Avarphi3}), (\ref{Ab1}), (\ref{Hx2}), and (\ref{abhat}) are employed.
Putting the above result into the first term in the second equality of
Eq.~(\ref{Eqbhat0})
leads to
\begin{eqnarray}
i{d\hat{b}_{\boldsymbol{k}\alpha_0}(t)\over dt}
&=&
[\hat{b}_{\boldsymbol{k}\alpha_0}(t),\hat{L}(t)]
\nonumber\\
&&-
\Big[\hat{b}_{\boldsymbol{k}\alpha_0}(t),
\sum_{\boldsymbol{k}\alpha_0^\prime \nu}
\hat{b}^\dagger_{\boldsymbol{k}\alpha_0^\prime}(t)
\hat{a}_{\boldsymbol{k}\nu}(t)
i{d \varphi^*_{\nu,\boldsymbol{k}\alpha_0^\prime}(t)\over d t}
\Big]
\nonumber\\
&&+
\sum_{\boldsymbol{k}\nu}
\hat{a}_{\boldsymbol{k}\nu}(t)i{d\varphi^*_{\nu,\boldsymbol{k}\alpha_0}(t)\over dt}
+i\hat{b}_{\boldsymbol{k}\alpha_0}(0)\delta(t)
\nonumber\\
&=&
[\hat{b}_{\boldsymbol{k}\alpha_0}(t),\hat{L}(t)]+i\hat{b}_{\boldsymbol{k}\alpha_0}(0)\delta(t),
\label{Eqbhat1}
\end{eqnarray}
where Eq.~(\ref{commbb}) is employed.
This becomes of the form of
\begin{equation}
i{d\hat{b}_{\boldsymbol{k}\alpha_0}(t)\over dt}=
E_{\alpha_0}(\boldsymbol{k})\hat{b}_{\boldsymbol{k}\alpha_0}(t)+i\hat{b}_{\boldsymbol{k}\alpha_0}(0)\delta(t),
\label{Eqbhat2}
\end{equation}
and the solution of it is given by
\begin{equation}
\hat{b}_{\boldsymbol{k}\alpha_0}(t)=\hat{b}_{\boldsymbol{k}\alpha_0}(0)e^{-iE_{\alpha_0}(\boldsymbol{k})t}\theta(t).
\end{equation}
Applying Eq.~(\ref{Ab1}), eventually, one obtains
\begin{equation}
\hat{b}_{\boldsymbol{k}\alpha}(t)=\hat{b}_{\boldsymbol{k}\alpha}(0)e^{-iE_{\alpha}(\boldsymbol{k})t}\theta(t),
\end{equation}
with $\alpha=\alpha_0(N_\alpha)$, and $\hat{b}_{\boldsymbol{k}\alpha_0}(0)=\hat{b}_{\boldsymbol{k}\alpha}(0)$.
Further, it should be noted that the equation of motion of $\hat{b}_{\boldsymbol{k}\alpha_0}(t)$ given by Eq.~(\ref{Eqbhat1}) for $t>0$ corresponds to the Heisenberg equation with the effective Hamiltonian $\hat{L}(t)$. 
This implies that the Floquet state $(\boldsymbol{k}\alpha_0)$ is considered as a fermionic quasiparticle, ensuring the anti-commutation relation Eq.~(\ref{commbb}).

According to the above results, the density matrix of Eq.~(\ref{rhoahat}) is recast into
\begin{eqnarray}
\rho_{\boldsymbol{k}\mu\nu}(t)=
\sum_{\alpha\alpha^\prime}
\bar{\rho}_{\boldsymbol{k}\alpha\alpha^\prime}(t)
C^*_{\mu 0,\boldsymbol{k}\alpha}C_{\nu 0,\boldsymbol{k}\alpha^\prime},
\label{rhoahat2}
\end{eqnarray}
where a density matrix of the quasiparticle is introduced as
\begin{eqnarray}
\bar{\rho}_{\boldsymbol{k}\alpha\alpha^\prime}(t)
&=&\langle \hat{b}^\dagger_{\boldsymbol{k}\alpha}(t)\hat{b}_{\boldsymbol{k}\alpha^\prime}(t) \rangle_g
\nonumber\\
&=&
\bar{\rho}_{\boldsymbol{k}\alpha\alpha^\prime}(0)
e^{i(E_\alpha(\boldsymbol{k})-E_{\alpha^\prime}(\boldsymbol{k}))t}\theta(t).
\label{rhobhat}
\end{eqnarray}
On the other hand, by employing Eq.~({\ref{bahat}}), one obtains
\begin{equation}
\bar{\rho}_{\boldsymbol{k}\alpha\alpha^\prime}(0)
=\sum_{\mu\nu nn^\prime}\rho_{\boldsymbol{k}\mu\nu}(0)
C_{\mu n,\boldsymbol{k}\alpha}C^*_{\nu n^\prime,\boldsymbol{k}\alpha^\prime},
\label{rhobhat2}
\end{equation}
providing the initial conditions of the quasiparticle states in Eq.~(\ref{rhobhat}) that are represented in terms of initial values of the original density matrix $\rho_{\boldsymbol{k}\mu\nu}(0)$.
Here, $\rho_{\boldsymbol{k}\mu\nu}(0)=\delta_{\mu\nu}\delta_{\mu v}$ with $v$ a band index corresponding to valence bands of states $|\Gamma_7, J_z=\pm 3/2\rangle$.

Moreover, by incorporating the phenomenological damping terms with the equation of motion of the density matrices $\rho_{\boldsymbol{k}\mu\nu}(t)$ of Eq.~(\ref{OBeqs}), $\bar{\rho}_{\boldsymbol{k}\alpha\alpha^\prime}(t)$ for $t > 0$ is modified to 
\begin{equation}
\bar{\rho}_{\boldsymbol{k}\alpha\alpha}(t)=\bar{\rho}_{\boldsymbol{k}\alpha\alpha}(0)e^{-t/T_1}+\bar{\rho}_{\boldsymbol{k}\alpha\alpha}(t_{qe})(1-e^{-t/T_1})
\label{OBED}
\end{equation}
for $\alpha =\alpha^\prime$, and
\begin{equation}
\bar{\rho}_{\boldsymbol{k}\alpha\alpha^\prime}(t)=\bar{\rho}_{\boldsymbol{k}\alpha\alpha^\prime}(0)
e^{i(E_\alpha(\boldsymbol{k})-E_{\alpha^\prime}(\boldsymbol{k}))t-t/T_2}
\label{OBEoffD}
\end{equation}
for $\alpha\not=\alpha^\prime$.\cite{SM1}
Here, $T_1$ and $T_2$ represent longitudinal and transverse relaxation times, respectively, and $t_{eq}$ represents quasi-equilibrium time, where $t_{eq}\gg T_1$.
Apparently, these expressions is in agreement with Eq.~(\ref{rhobhat}) in the limit of $T_1,\:T_2\rightarrow \infty$.

\subsection{Laser induced magnetization}
\label{sec2C}

Here, one presents an exact quantum-mechanical description of an optically induced magnetization imparted by an intense laser with circular polarization within the electric dipole approximation.
This provides a materials-specific framework for the calculation
of the optomagnetic polarization in a non-perturbative manner with respect to laser intensity, not limited to model systems.

\subsubsection{Solenoid current density and magnetization}
\label{sec2C1}

One begins with the following Hamiltonian as
\begin{equation}
h(\boldsymbol{r},t)=h^0(\boldsymbol{r})+V(\boldsymbol{r},t),
\label{h}
\end{equation}
where $h^0(\boldsymbol{r})$ represents a one-electron material Hamiltonian, given by
\begin{equation}
h^0(\boldsymbol{r},t)
=
-{\boldsymbol{\nabla}^2\over 2}+v(\boldsymbol{r})+v_{so}(\boldsymbol{r}),
\label{h0}
\end{equation}
with $v(\boldsymbol{r})$ and $v_{so}(\boldsymbol{r})$ as a periodic potential of crystal and a spin-orbit interaction, respectively, and $V(\boldsymbol{r},t)$ represents the electric dipole interaction with the laser, given by
\begin{equation}
V(\boldsymbol{r},t)=\boldsymbol{F}(t)\cdot\boldsymbol{r}.
\label{V}
\end{equation}
Here, a field operator $\hat{\Psi}(\boldsymbol{r},t)$ is defined so as to ensure the equation of motion
\begin{equation}
i{\partial \hat{\Psi}(\boldsymbol{r},t)\over \partial t }
=h(\boldsymbol{r},t)\hat{\Psi}(\boldsymbol{r},t),
\label{EqPsiprime}
\end{equation}
with the equal-time anti-commutation relation
\begin{equation}
\{
\hat{\Psi}(\boldsymbol{r},t), \hat{\Psi}^{\dagger}(\boldsymbol{r}^\prime,t)
\}
=\delta(\boldsymbol{r}-\boldsymbol{r}^\prime).
\end{equation}
Accordingly, it is readily seen that the charge density of electron represented by
\begin{equation}
\rho(\boldsymbol{r},t)
=-\langle \hat{\Psi}^\dagger(\boldsymbol{r},t)\hat{\Psi}(\boldsymbol{r},t)\rangle_g
\end{equation}
follows the equation of continuity:
\begin{equation}
{\partial \rho(\boldsymbol{r},t)\over \partial t}
+\boldsymbol{\nabla}\cdot\boldsymbol{j}(\boldsymbol{r},t)=0,
\label{cont}
\end{equation}
where $\boldsymbol{j}(\boldsymbol{r},t)$ represents the charge current density of electron, given by
\begin{equation}
\boldsymbol{j}(\boldsymbol{r},t)
=
-\Re\langle
-i\hat{\Psi}^{\dagger}(\boldsymbol{r},t)\boldsymbol{\nabla}\hat{\Psi}(\boldsymbol{r},t)
\rangle_g.
\label{j}
\end{equation}

On the other hand, Amp\`{e}re's circuital law is read as
\begin{equation}
\boldsymbol{\nabla}\times{\boldsymbol{B}(\boldsymbol{r},t)\over \mu_0}
=\boldsymbol{j}_{tot}(\boldsymbol{r},t)
+{\partial \varepsilon_0\boldsymbol{E}(\boldsymbol{r},t)\over \partial t},
\label{amp}
\end{equation}
where $\boldsymbol{E}(\boldsymbol{r},t)$ and $\boldsymbol{B}(\boldsymbol{r},t)$ represent an electric field and a magnetic flux density, respectively, with $\varepsilon_0$ and $\mu_0$ the permittivity of vacuum and the permeability of vacuum, respectively.
The total current density $\boldsymbol{j}_{tot}(\boldsymbol{r},t)$ is provided by the sum of the free current density $\boldsymbol{j}(\boldsymbol{r},t)$ and bound current densities as
\begin{equation}
\boldsymbol{j}_{tot}(\boldsymbol{r},t)
=\boldsymbol{j}(\boldsymbol{r},t)+\boldsymbol{j}_m(\boldsymbol{r},t)+\boldsymbol{j}_p(\boldsymbol{r},t),
\label{jtot0}
\end{equation}
where $\boldsymbol{j}_m(\boldsymbol{r},t)$ and $\boldsymbol{j}_p(\boldsymbol{r},t)$ represent the bound currents ascribable to magnetization $\boldsymbol{M}(\boldsymbol{r},t)$ and polarization $\boldsymbol{P}(\boldsymbol{r},t)$, given by
\begin{equation}
\boldsymbol{j}_m(\boldsymbol{r},t)
=\boldsymbol{\nabla}\times\boldsymbol{M}(\boldsymbol{r},t),
\label{jm}
\end{equation}
and
\begin{equation}
\boldsymbol{j}_p(\boldsymbol{r},t)
={\partial \boldsymbol{P}(\boldsymbol{r},t)\over \partial t},
\label{jp}
\end{equation}
respectively.
Since the polarization vector is read as
\begin{equation}
\boldsymbol{P}(\boldsymbol{r},t)
=-\langle \hat{\Psi}^\dagger(\boldsymbol{r},t)\boldsymbol{r}\hat{\Psi}(\boldsymbol{r},t)\rangle_g
=\boldsymbol{r}\rho(\boldsymbol{r},t),
\end{equation}
the bound current density induced by this is rewritten as
\begin{equation}
\boldsymbol{j}_p(\boldsymbol{r},t)
=\boldsymbol{r}{\partial \rho(\boldsymbol{r},t)\over \partial t}
=-\boldsymbol{r} \boldsymbol{\nabla}\cdot \boldsymbol{j}(\boldsymbol{r},t),
\label{jp2}
\end{equation}
due to Eq.~(\ref{cont}).

Employing the following formula\cite{Hertel2006}
\begin{equation}
\boldsymbol{A}\boldsymbol{\nabla}\cdot\boldsymbol{B}
-\boldsymbol{B}\boldsymbol{\nabla}\cdot\boldsymbol{A}
=\boldsymbol{\nabla}\times(\boldsymbol{A}\times\boldsymbol{B})
+\boldsymbol{A}\cdot\boldsymbol{\nabla}\boldsymbol{B}
-\boldsymbol{B}\cdot\boldsymbol{\nabla}\boldsymbol{A}
\end{equation}
with $\boldsymbol{A}=-\boldsymbol{r}$ and $\boldsymbol{B}=\boldsymbol{j}(\boldsymbol{r},t)$, Eq.~(\ref{jp2}) becomes
\begin{eqnarray}
\boldsymbol{j}_p(\boldsymbol{r},t)
&=&
-\boldsymbol{\nabla}\times[\boldsymbol{r}\times\boldsymbol{j}(\boldsymbol{r},t)]
-\boldsymbol{r}\cdot\boldsymbol{\nabla}\boldsymbol{j}(\boldsymbol{r},t)
\nonumber\\
&&
-2\boldsymbol{j}(\boldsymbol{r},t).
\end{eqnarray}
This brings the free current density to be of the form
\begin{eqnarray}
\boldsymbol{j}(\boldsymbol{r},t)
&=&-{1\over 2}\Big\{
\boldsymbol{\nabla}\times[\boldsymbol{r}\times\boldsymbol{j}(\boldsymbol{r},t)]
+\boldsymbol{r}\cdot\boldsymbol{\nabla}\boldsymbol{j}(\boldsymbol{r},t)
\nonumber\\
&&
-\boldsymbol{r} \boldsymbol{\nabla}\cdot \boldsymbol{j}(\boldsymbol{r},t)
\Big\}.
\end{eqnarray}
Therefore, the total current density becomes
\begin{eqnarray}
\boldsymbol{j}_{tot}(\boldsymbol{r},t)
&=&\boldsymbol{j}(\boldsymbol{r},t)+\boldsymbol{\nabla}\times\boldsymbol{M}(\boldsymbol{r},t)-\boldsymbol{r} \boldsymbol{\nabla}\cdot \boldsymbol{j}(\boldsymbol{r},t)
\nonumber\\
&=&\boldsymbol{\nabla}\times
\left\{\boldsymbol{M}(\boldsymbol{r},t)
-{1\over 2}
\boldsymbol{r}\times\boldsymbol{j}(\boldsymbol{r},t)\right\}
\nonumber\\
&&
-{1\over 2}
\left\{
\boldsymbol{r}\cdot\boldsymbol{\nabla}\boldsymbol{j}(\boldsymbol{r},t)
+\boldsymbol{r} \boldsymbol{\nabla}\cdot \boldsymbol{j}(\boldsymbol{r},t)
\right\}.
\label{jtot2}
\end{eqnarray}
It is obvious that the first term in the last equality implies the solenoid current density $\boldsymbol{j}^{(sol)}(\boldsymbol{r},t)$ imparted by $\boldsymbol{j}_{tot}(\boldsymbol{r},t)$.
This gives rise to the magnetization given by
\begin{equation}
\boldsymbol{M}^{(sol)}(\boldsymbol{r},t)=\boldsymbol{M}(\boldsymbol{r},t)
-{1\over 2}
\boldsymbol{r}\times\boldsymbol{j}(\boldsymbol{r},t).
\label{Msol}
\end{equation}
Here, the first term originates simply from spins of electrons, and the other is ascribable to the polarization-induced magnetization.

The first term of Eq.~(\ref{Msol}) is expressed as
\begin{equation}
\boldsymbol{M}(\boldsymbol{r},t)=\mu_Bg_e\langle \hat{\Psi}^\dagger(\boldsymbol{r},t)\boldsymbol{s}\hat{\Psi}(\boldsymbol{r},t)\rangle_g
\label{M}
\end{equation}
with the spin angular momentum operator as $\boldsymbol{s}=\boldsymbol{\sigma}/2$,
where
Bohr magneton $\mu_B$ is introduced, namely,
\(
\mu_B=1/2,
\)
and $g_e$ represents the gyromagnetic ratio of electron spin.
Moreover, the second term is given by
\begin{equation}
\boldsymbol{M}^{(2)}(\boldsymbol{r},t)
=\mu_B
\Re\left[
\langle\hat{\Psi}^\dagger(\boldsymbol{r},t)\boldsymbol{l}\hat{\Psi}(\boldsymbol{r},t)\rangle_g
\right].
\label{M2}
\end{equation}
Here, this term ascribable to $\boldsymbol{l}$ shows the contribution of the orbital magnetization caused by an elliptic trajectory of electron under the circularly polarized light field.
The effect of magnetization thus induced is termed as the IFE as mentioned in Sec.~\ref{sec1}, which is understood as a second-order optical process caused exclusively by the irradiation of circularly polarized light.
The above procedure of extracting the contribution of the orbital magnetization from $\boldsymbol{j}^{(sol)}(\boldsymbol{r},t)$ is also adopted in the perturbative manner in Ref.~\onlinecite{Hertel2006}.
The obtained expression of IFE in terms of $\boldsymbol{l}$ is identical with that introduced in Ref.~\onlinecite{Battiato2014}, however, without a quantum-mechanical derivation.
As shown right below, in fact, the term of $\boldsymbol{l}$ is altered by the choice of a gauge factor incorporated in $\hat{\Psi}(\boldsymbol{r},t)$ as an additional phase.
Equation~(\ref{M2}) is applicable to calculations of the optomagnetic polarization of any material induced by a circularly polarized laser with any laser intensity regardless of temporal width---ranging from an ultrashort pulse to a continuous wave laser.
To be more specific, this is evaluated in terms of transient density matrices subject to non-equilibrium quantum dynamics, as shown below, differing from most of theories of IFE just providing a stationary part of induced magnetization at an asymptotic limit.\cite{Pershan1966,Hertel2006,Zhang2009,Woodford2009,Kurkin2008,Taguchi2011,Popova2011,Popova2012,Gridnev2013,Qaiumzadeh2013}

In passing, as regards the second term of Eq.~(\ref{jtot2}), it is shown that
\begin{equation}
\boldsymbol{\nabla}\cdot \boldsymbol{j}_{tot}(\boldsymbol{r},t)=
-{1\over 2}\boldsymbol{\nabla}\cdot
\left\{
\boldsymbol{r}\cdot\boldsymbol{\nabla}\boldsymbol{j}(\boldsymbol{r},t)
+\boldsymbol{r} \boldsymbol{\nabla}\cdot \boldsymbol{j}(\boldsymbol{r},t)
\right\},
\label{jtot}
\end{equation}
while owing to Gauss' law, Eq.~(\ref{amp}) becomes
\begin{eqnarray}
0&=&\boldsymbol{\nabla}\cdot \boldsymbol{j}_{tot}(\boldsymbol{r},t)+
{\partial \varepsilon_0\boldsymbol{\nabla}\cdot\boldsymbol{E}(\boldsymbol{r},t)\over \partial t}
\nonumber\\
&=&
\boldsymbol{\nabla}\cdot \boldsymbol{j}_{tot}(\boldsymbol{r},t)+
{\partial \rho_{tot}(\boldsymbol{r},t)\over \partial t},
\label{cont3}
\end{eqnarray}
where $\rho_{tot}(\boldsymbol{r},t)$ represents the total charge density defined by the sum of
$\rho(\boldsymbol{r},t)$ and the polarized charge 
$\rho_p(\boldsymbol{r},t)=-\boldsymbol{\nabla}\cdot\boldsymbol{P}(\boldsymbol{r},t)$, that is,
\begin{equation}
\rho_{tot}(\boldsymbol{r},t)=\rho(\boldsymbol{r},t)+\rho_{p}(\boldsymbol{r},t).
\end{equation}
Hence, Eq.~(\ref{jtot}) is nothing but the equation of continuity for the total charge density and current density, and the right side of Eq.~(\ref{jtot}) is identical with $-\partial \rho_{tot}(\boldsymbol{r},t)/\partial t$.

\subsubsection{Gauge transformation}
\label{sec2C2}

Below, we aim to represent Eqs.~(\ref{M}) and (\ref{M2}) in terms of the density matrix of Eq.~(\ref{rhoahat}) by taking an appropriate gauge transformation on $\hat{\Psi}(\boldsymbol{r},t)$.
Hereafter, a function $f(\boldsymbol{r})$ and an operator $\hat{\xi}(\boldsymbol{r})$ are represented as $\langle \boldsymbol{r}|f\rangle$ and $\langle \boldsymbol{r}|\hat{\xi}\rangle $, respectively, in terms of the Dirac notation, if necessary.
Accordingly, Eq.~(\ref{h}) is rewritten as
\begin{equation}
h(\boldsymbol{r},t)=\sum_{\mu,\boldsymbol{R},\mu^\prime,\boldsymbol{R}^\prime}
\langle\boldsymbol{r}|\mu\boldsymbol{R}\rangle h_{\mu\boldsymbol{R},\mu^\prime \boldsymbol{R}^\prime}(t)\langle \mu^\prime\boldsymbol{R}^\prime|\boldsymbol{r}\rangle
\end{equation}
in terms of a set of Wannier states $\{|\mu\boldsymbol{R}\rangle\}$ with $\mu$ and $\boldsymbol{R}$ a band index and a site vector of the crystal.
Here, 
\begin{eqnarray}
h_{\mu\boldsymbol{R},\mu^\prime \boldsymbol{R}^\prime}(t)
&=&\langle \mu\boldsymbol{R}|h^0|\mu^\prime \boldsymbol{R}^\prime\rangle
+\langle \mu\boldsymbol{R}|V|\mu^\prime \boldsymbol{R}^\prime\rangle
\nonumber\\
&\equiv&
h^0_{\mu\boldsymbol{R},\mu^\prime \boldsymbol{R}^\prime}+
V_{\mu\boldsymbol{R},\mu^\prime \boldsymbol{R}^\prime}(t).
\end{eqnarray}
The second term is represented by
\begin{eqnarray}
V_{\mu\boldsymbol{R},\mu^\prime \boldsymbol{R}^\prime}(t)
&=&\langle \mu\boldsymbol{R}|\boldsymbol{F}(t)\cdot \boldsymbol{r}|\mu^\prime\boldsymbol{R}^\prime\rangle
\nonumber\\&=&
\boldsymbol{F}(t)\cdot \boldsymbol{R}\:\delta_{\mu\mu^\prime}\delta_{\boldsymbol{R}\boldsymbol{R}^\prime}
\nonumber\\
&&
+\boldsymbol{F}(t)\cdot \boldsymbol{X}_{\mu\mu^\prime}\bar{\delta}_{\mu\mu^\prime}\delta_{\boldsymbol{R}\boldsymbol{R}^\prime},
\end{eqnarray}
where $\boldsymbol{X}$ is defined as the displacement of electron from the position of an ionic core of the crystal, namely, $\boldsymbol{X}=\boldsymbol{r}-\boldsymbol{R}$,
and an electric dipole matrix element at each $\boldsymbol{R}$ is approximately given by
\(
\langle \mu\boldsymbol{R}|\boldsymbol{X}|\mu^\prime\boldsymbol{R}^\prime\rangle
\approx \boldsymbol{r}_{\mu\mu^\prime}\delta_{\boldsymbol{R}\boldsymbol{R}^\prime} 
\)
with 
\(
\boldsymbol{r}_{\mu\mu^\prime}=\langle \mu|\boldsymbol{r}|\mu^\prime\rangle.
\)
Here, the property has been used that a Wannier function $\langle \boldsymbol{r}|\mu\boldsymbol{R}\rangle$ is just a function of $\boldsymbol{X}$, and is mostly localized inside a cell situated at $\boldsymbol{R}$.
Thus, $\langle \mu\boldsymbol{R}|\boldsymbol{X}|\mu^\prime\boldsymbol{R}^\prime\rangle$ is considered almost independent
of $\boldsymbol{R}$.
Therefore, Eq.~(\ref{EqPsiprime}) is cast into
\begin{eqnarray}
i{\partial\over \partial t}\langle \mu\boldsymbol{R}|\hat{\Psi}\rangle
&=&
\sum_{\mu^\prime\boldsymbol{R}^\prime}
\left\{
h^0_{\mu\boldsymbol{R},\mu^\prime\boldsymbol{R}^\prime}
+\boldsymbol{F}(t)\cdot(\boldsymbol{R}\delta_{\mu\mu^\prime}
\right.
\nonumber\\
&&
\left.
+\boldsymbol{r}_{\mu\mu^\prime}\bar{\delta}_{\mu\mu^\prime})
\delta_{\boldsymbol{R}\boldsymbol{R}\boldsymbol{}^\prime}
\right\}
\langle \mu^\prime\boldsymbol{R}^\prime|\hat{\Psi} \rangle.
\label{H2}
\end{eqnarray}
Now, the Peierls phase transform defined by
\begin{equation}
\langle \mu\boldsymbol{R}|\hat{\Psi} \rangle=e^{i\boldsymbol{A}(t)\cdot\boldsymbol{R}}
\langle \mu\boldsymbol{R}|\hat{\Phi} \rangle
\label{peierls}
\end{equation}
brings Eq.~(\ref{H2}) into the following equation for 
$\hat{\Phi}_{\mu\boldsymbol{R}} \equiv \langle \mu\boldsymbol{R}|\hat{\Phi}\rangle$ as
\begin{equation}
\sum_{\mu^\prime\boldsymbol{R}^\prime}
e^{-i\boldsymbol{A}(t)\cdot\boldsymbol{R}}\tilde{h}_{\mu\boldsymbol{R},\mu^\prime\boldsymbol{R}^\prime}(t)
e^{i\boldsymbol{A}(t)\cdot\boldsymbol{R}^\prime}\hat{\Phi}_{\mu^\prime\boldsymbol{R}^\prime}
=i{\partial\over \partial t}\hat{\Phi}_{\mu\boldsymbol{R}},
\label{H3}
\end{equation}
where
\begin{eqnarray}
\tilde{h}_{\mu\boldsymbol{R},\mu^\prime\boldsymbol{R}^\prime}(t)
&\equiv&
\langle \mu\boldsymbol{R}|\tilde{h}(t)|\mu^\prime\boldsymbol{R}^\prime\rangle
\nonumber\\
&=&
h^0_{\mu\boldsymbol{R},\mu^\prime\boldsymbol{R}^\prime}
+\boldsymbol{F}(t)\cdot \boldsymbol{r}_{\mu\mu^\prime}\bar{\delta}_{\mu\mu^\prime}\delta_{\boldsymbol{R}\boldsymbol{R}}.
\end{eqnarray}
Here, $\tilde{h}(t)$ is defined by
\begin{eqnarray}
\tilde{h}(\boldsymbol{r},\boldsymbol{r}^\prime;t)
&=&\langle\boldsymbol{r}|\tilde{h}(t)|\boldsymbol{r}^\prime\rangle
\nonumber\\
&=&
\delta(\boldsymbol{r}-\boldsymbol{r}^\prime)h^0(\boldsymbol{r})+\boldsymbol{F}(t)\cdot \boldsymbol{\chi}(\boldsymbol{r},\boldsymbol{r}^\prime),
\label{tildeh}
\end{eqnarray}
where
\begin{equation}
\boldsymbol{\chi}(\boldsymbol{r},\boldsymbol{r}^\prime)
=
\sum_{\mu\mu^\prime\boldsymbol{R}}
\langle\boldsymbol{r}|\mu\boldsymbol{R}\rangle
\boldsymbol{r}_{\mu\mu^\prime}\bar{\delta}_{\mu\mu^\prime}
\langle \mu^\prime\boldsymbol{R}|\boldsymbol{r}^\prime\rangle.
\end{equation}

The following equation as
\begin{equation}
\hat{\Phi}_{\mu\boldsymbol{k}}(t)
\equiv\langle \mu\boldsymbol{k}|\hat{\Phi}(t) \rangle
=\hat{a}_{\boldsymbol{k}\mu}(t)
\label{Phia}
\end{equation}
is derived from Eq.~(\ref{H3}), as shown in Appendix~\ref{app2}, where
$| \mu\boldsymbol{k}\rangle$ represents a Bloch state.
Thus, by taking account of Eq.~(\ref{peierls}), $\hat{\Psi}(\boldsymbol{r},t)=\langle \boldsymbol{r}|\hat{\Psi}(t)\rangle$ becomes of the form:
\begin{eqnarray}
\hat{\Psi}(\boldsymbol{r},t)
&=&
{1\over \sqrt{N}}
\sum_{\mu\boldsymbol{R}\boldsymbol{k}}
\langle \boldsymbol{r}|\mu\boldsymbol{R}\rangle e^{i{(\boldsymbol{k}+\boldsymbol{A}(t))\cdot\boldsymbol{R}}}\:\hat{a}_{\boldsymbol{k}\mu}(t)
\nonumber\\
&=&
\sum_{\mu\boldsymbol{k}}
e^{i{(\boldsymbol{k}+\boldsymbol{A}(t))\cdot\boldsymbol{r}}}
U_{\mu(\boldsymbol{k}+\boldsymbol{A}(t))}(\boldsymbol{r})\:\hat{a}_{\boldsymbol{k}\mu}(t)
\nonumber\\
&\approx&
\sum_{\mu\boldsymbol{k}}
e^{i{(\boldsymbol{k}+\boldsymbol{A}(t))\cdot\boldsymbol{r}}}
U_{\mu\boldsymbol{0}}(\boldsymbol{r})\hat{a}_{\boldsymbol{k}\mu}(t)
\nonumber\\
&=&
{1\over \sqrt{N}}
\sum_{\mu\boldsymbol{R}\boldsymbol{k}}
e^{i{(\boldsymbol{k}+\boldsymbol{A}(t))\cdot\boldsymbol{r}}}
\langle \boldsymbol{r}-\boldsymbol{R}|\chi_\mu\rangle
\langle \chi_\mu|\hat{\psi}_{\boldsymbol{k}}(t)\rangle
\nonumber\\
&=&
{1\over \sqrt{N}}
\sum_{\boldsymbol{R}\boldsymbol{k}}
e^{i{(\boldsymbol{k}+\boldsymbol{A}(t))\cdot\boldsymbol{r}}}
\langle \boldsymbol{r}-\boldsymbol{R}|\hat{\psi}_{\boldsymbol{k}}(t)\rangle.
\label{Psix}
\end{eqnarray}
Here, in the second equality, a periodic part of the Bloch function $\langle \boldsymbol{r}|\mu\boldsymbol{k}\rangle$ is introduced as $U_{\mu\boldsymbol{k}}(\boldsymbol{r})$, in the third equality, the approximation of $U_{\mu\boldsymbol{K}(t)}(\boldsymbol{r})\approx U_{\mu\boldsymbol{0}}(\boldsymbol{r})$ is made to be consistent with the spirit of the $\boldsymbol{k}\cdot\boldsymbol{p}$ model employed in Eq.~(\ref{kp}), and in the fourth equality, Eqs.~(\ref{Psihat}) and (\ref{U0}) are employed and the Wannier function $\langle \boldsymbol{r}|\mu\boldsymbol{R}\rangle$ is represented as $\langle \boldsymbol{r}-\boldsymbol{R}|\chi_\mu\rangle\equiv \chi_\mu(\boldsymbol{r}-\boldsymbol{R})$.

It is evident that the gauge transformation
\begin{equation}
\hat{\Psi}(\boldsymbol{r},t)=e^{i\boldsymbol{A}(t)\cdot \boldsymbol{r}}\hat{\Psi}^\prime(\boldsymbol{r},t)
\end{equation}
of Eq.~(\ref{Psix}) is attributed to the Peierls phase transform of Eq.~(\ref{peierls}),
where
\begin{equation}
\hat{\Psi}^\prime(\boldsymbol{r},t)
=
{1\over \sqrt{N}}
\sum_{\boldsymbol{R}\boldsymbol{k}} e^{i{\boldsymbol{k}\cdot\boldsymbol{r}}}
\hat{\psi}_{\boldsymbol{k}}(\boldsymbol{r}-\boldsymbol{R},t).
\end{equation}
Applying this to Eqs.~(\ref{M}) and (\ref{M2}), these become
\begin{equation}
\boldsymbol{M}(\boldsymbol{r},t)
=
\mu_Bg_e\langle \hat{\Psi}^{\prime\dagger}(\boldsymbol{r},t)\boldsymbol{s}\hat{\Psi}^\prime(\boldsymbol{r},t)\rangle_g,
\label{Mprime}
\end{equation}
and
\begin{eqnarray}
&&\boldsymbol{M}^{(2)}(\boldsymbol{r},t)
\nonumber\\
&&
=
\mu_B
\Re\left[
\langle\hat{\Psi}^{\prime\dagger}(\boldsymbol{r},t)\left\{\boldsymbol{l}
+\boldsymbol{r}\times\boldsymbol{A}(t)\right\}\hat{\Psi}^\prime(\boldsymbol{r},t)\rangle_g
\right],
\label{M2prime}
\end{eqnarray}
respectively.
It is seen in Eq.~(\ref{M2prime}) that the operator $\boldsymbol{l}$ for the IFE is altered by an additional contribution depending on $\boldsymbol{A}(t)$, though the physical quantity $\boldsymbol{M}^{(2)}(\boldsymbol{r},t)$ remains invariant with respect to this gauge transformation.

Finally, the volume-averaged magnetization at time $t$
\begin{equation}
\bar{\boldsymbol{M}}^{(sol)}(t)
=\bar{\boldsymbol{M}}(t)+\bar{\boldsymbol{M}}^{(2)}(t),
\label{Msol2}
\end{equation}
is shown below, where
\begin{equation}
\bar{\mathcal{\boldsymbol{M}}}(t)
=
{1\over V}\int 
d\boldsymbol{r}\mathcal{\boldsymbol{M}}(\boldsymbol{r},t)
\end{equation}
with
$\mathcal{\boldsymbol{M}}=\boldsymbol{M}, \boldsymbol{M}^{(2)}$.
Equation~(\ref{Mprime}) becomes
\begin{eqnarray}
\bar{\boldsymbol{M}}(t)
&=&
{\mu_Bg_e\over V}
\sum_{\boldsymbol{k}\mu\nu\Delta\boldsymbol{R}}
\int d \boldsymbol{X} \chi^*_\mu (\boldsymbol{X}) 
\boldsymbol{s} 
\chi_\nu(\boldsymbol{X}+\Delta\boldsymbol{R})
\nonumber\\
&&\times
\langle a^\dagger_{\boldsymbol{k}\mu}(t)a_{\boldsymbol{k}\nu}(t)\rangle_g
\nonumber\\
&\approx&
{\mu_Bg_e\over V}
\sum_{\boldsymbol{k}\mu\nu}
\rho_{\boldsymbol{k}\mu\nu}(t)
\langle\mu|\boldsymbol{s}|\nu\rangle,
\label{Mbar}
\end{eqnarray}
where
in the last equality, just the Wannier functions at the same site, namely, $\Delta\boldsymbol{R}=\boldsymbol{0}$ are taken account of for the summation over $\Delta\boldsymbol{R}$, and the density matrix of Eq.~(\ref{rhoahat}) is introduced.
Similarly, Eq.~(\ref{M2prime}) ends up with
\begin{eqnarray}
\bar{\boldsymbol{M}}^{(2)}(t)
&=&
{\mu_B\over V}\sum_{\boldsymbol{k}\mu\nu}
\Re \Big[\rho_{\boldsymbol{k}\mu\nu}(t)
\Big\{\langle \mu|\boldsymbol{l}|\nu\rangle
\nonumber\\
&&
\hspace{1cm}
+
\langle \mu|\boldsymbol{r}|\nu\rangle
\times\boldsymbol{K}(t)
\Big\}
\Big]
\label{Mbar2}
\end{eqnarray}
with $\boldsymbol{K}(t)=\boldsymbol{k}+\boldsymbol{A}(t)$.
By numerically evaluating $\bar{\boldsymbol{M}}^{(sol)}(t)$ at each $t$, the temporal variance of it is examined, followed by extracting the time-independent magnetization at the asymptotic limit, $\bar{\boldsymbol{M}}^{(sol)}(t\rightarrow\infty)$.
In actual calculations, the expression of Eq.~(\ref{rhoahat2}) for the density matrices is employed in light of of Eqs.~(\ref{OBED}) and (\ref{OBEoffD}).

\section{Results and Discussion}
\label{sec3}

\subsection{Floquet energy bands and Weyl node formation}
\label{sec3A}

For the purpose of the later discussion of induced magnetization, one shows the band structures of Floquet energy $E_\alpha(\boldsymbol{k})$ that are calculated based on the theoretical framework of Sec.~\ref{sec2A}.
For the numerical calculations, the same material parameters as those of Ref.~\onlinecite{Zhang2022} are employed; the bandgap of the original crystal Zn$_3$As$_2$ is set to be $E_g=$0.0169 (0.46 eV).
Hereafter, the conduction bands with $J_z=\pm 1/2$ and the heavy-hole bands with $J_z=\pm 3/2$ are denoted as $e$ and $hh$, respectively, and thus the associated Floquet sidebands with $N$ photons are represented by $e(N)$ and $hh(N)$, respectively.
In particular, an attention is paid to the Floquet states of
$e(0), e(-1), hh(0)$, and $hh(+1)$ around the Fermi energy $E_F=0$, which is set at the middle of the energy difference between the original $e$ and $hh$ bands at the $\Gamma$-point.
It is likely that the band of $e(-1)$ crosses with that of $hh(+1)$ in a certain range of laser frequency $\omega$ and strength of electric field $F_c$ to form a pair of Weyl nodes on the $k_z$ axis, accompanying band deformation.
Strictly speaking, pairs of Weyl nodes manifest themselves in many places in the BK, at which the Floquet bands $e(N)$ and $hh(N^\prime)$ intersect regardless of the magnitude of $\omega(\not=0)$; for more detail, consult Fig.~6 of Ref~\onlinecite{Zhang2022}.

\begin{figure}[t]
\begin{center}
\includegraphics[width=8.5cm]{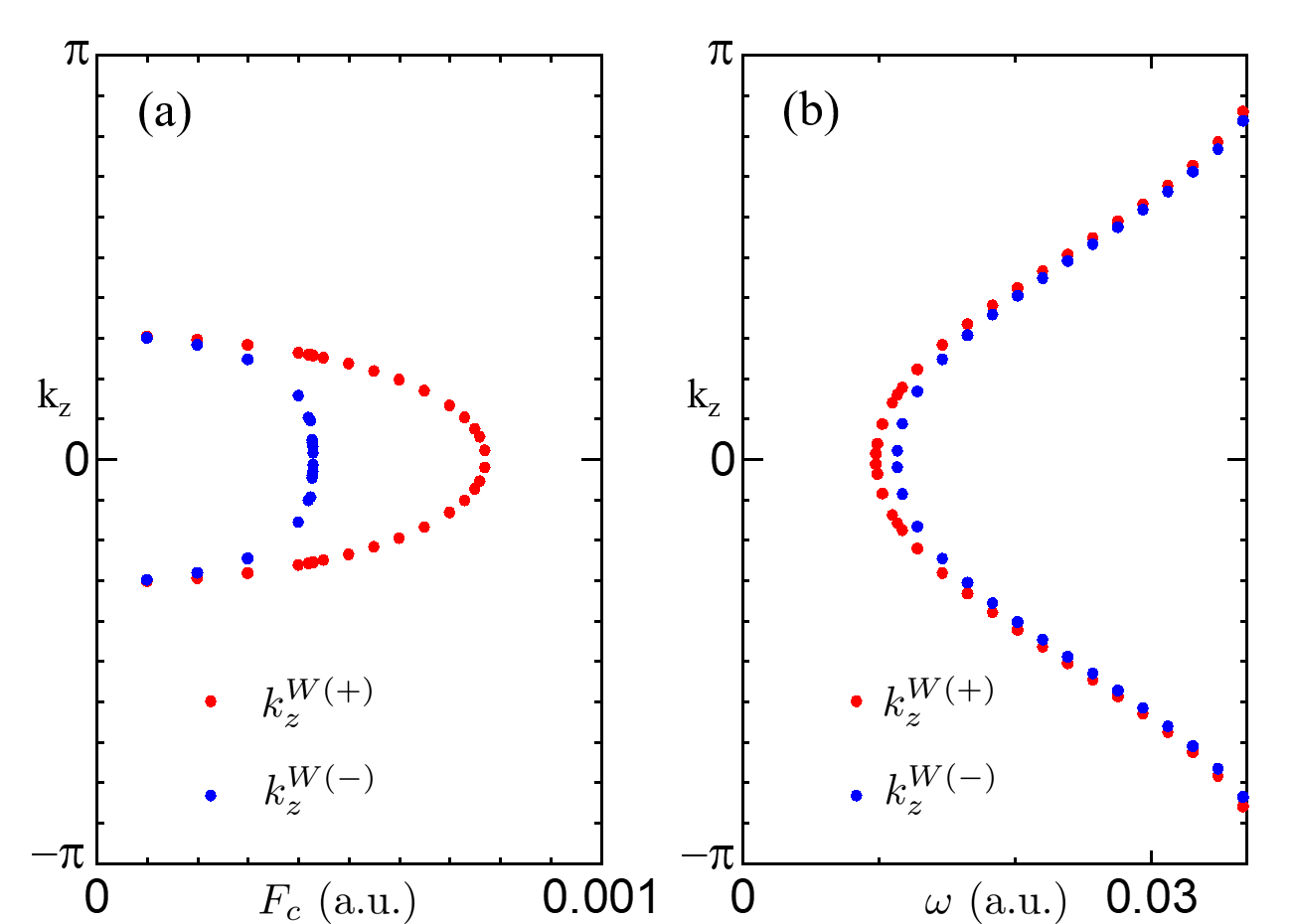}
\end{center}
\caption{Variance of locations of the Weyl nodes at $k_z^{W(+)}$ and $k_z^{W(-)}$ on the $k_z$ axis for up-spin (denoted by red dots) and down-spin (denoted by blue dots) states of FWSM, respectively.
(a) $k_z^{W(\pm)}$ as a function of $F_c$ with $\omega=$0.0147 (0.4 eV).
The axis of ordinate is gauged in the unit of $1/d_z$.
(b) The same as panel (a) but as a function of $\omega$ with $F_c=$0.0003 (1.54 MV/cm).
}
\label{fig1}
\end{figure}

Figure~\ref{fig1}(a) shows the variance of locations of the Weyl nodes at $k_z^{W(+)}$ and $k_z^{W(-)}$ on the $k_z$ axis for up-spin and down-spin states of FWSM, respectively, as a function of $F_c$ with $\omega=$0.0147 (0.4 eV).
Figure~\ref{fig1}(b) shows the same as Fig.~\ref{fig1}(a) but as a function of $\omega$ with $F_c=$0.0003 (1.54 MV/cm).
It is found that one Weyl node at $k_z^{W(\pm)}$ is annihilated with another Weyl node formed at $-k_z^{W(\pm)}$ with an opposite chirality at $k_z^{W(\pm)}=0$ where these are brought together.
This shows a topological phase transition to a trivial insulating phase with the opening of a Floquet band gap.
Another topological phase transition to a non-trivial insulating phase at $k_z^{W(\pm)}=\pi$ would be likely caused for $\omega > E_g$ as shown in Fig.~\ref{fig1}(b), while this is not verified in Fig.~\ref{fig1}(a); where $k_z^{W(\pm)}=0.95$ at $F_c \rightarrow 0$ following an analytic model derived in Ref.~\onlinecite{Zhang2022}.
The variance of locations of the Weyl nodes is consulted in Sec.~\ref{sec3B} for the discussion of the $F_c$ and $\omega$ dependence of the induced magnetization.

\begin{figure}[t]
\begin{center}
\includegraphics[width=8.5cm]{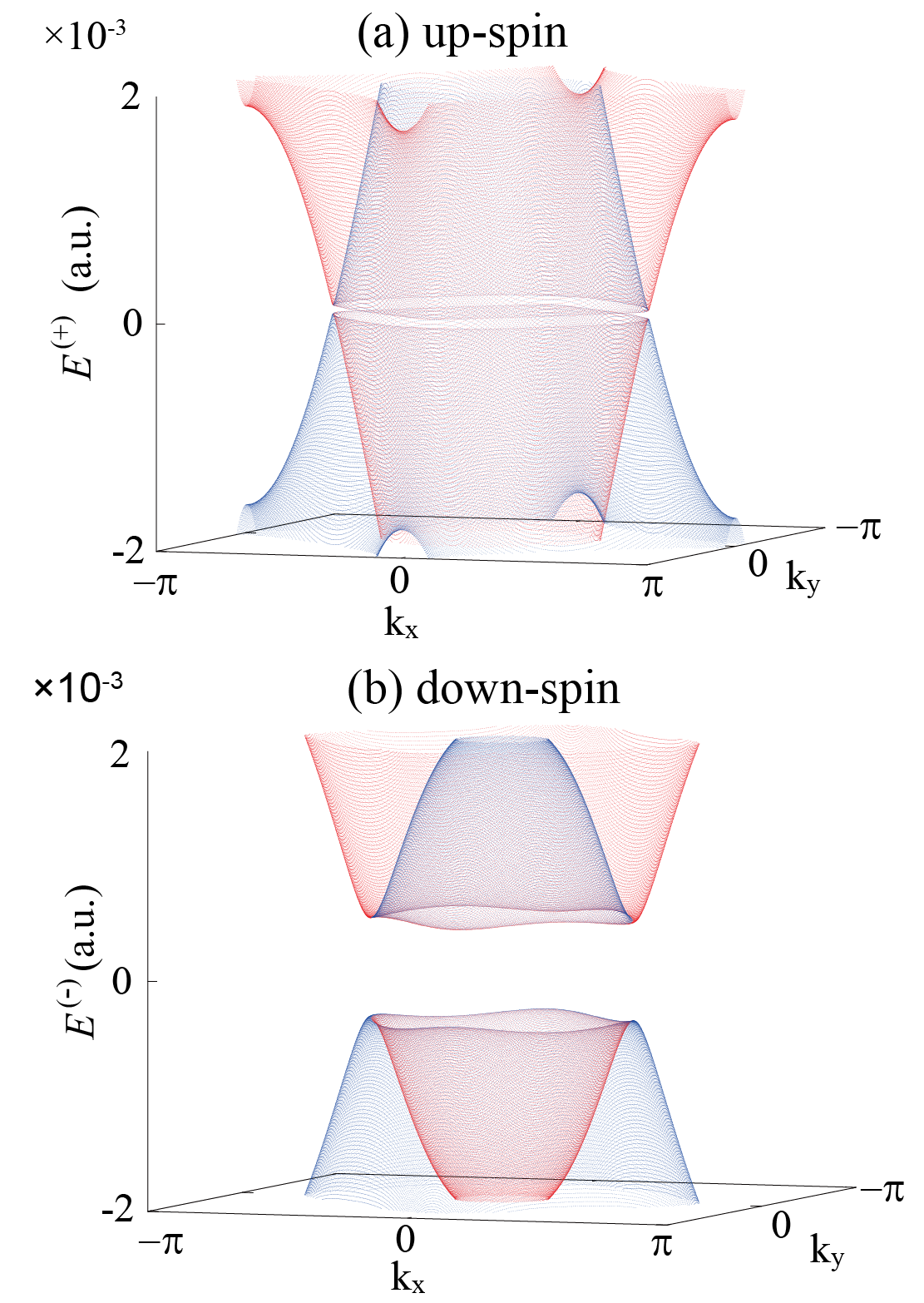}
\end{center}
\caption{
Bulk energy dispersion of $E^{(\pm)}(\boldsymbol{k})$ of FWSM bands of $e(-1)$ and $hh(+1)$ at $k_z = 0$ in the atomic units, generated by the laser drive with circular polarization, where $F_c=0.0003$ (1.54 MV/cm), and $\omega=$0.0147 (0.4 eV). 
Here, bands dominated rather by the $s$-orbital ($p$-orbital) component are denoted by a red (blue) solid line. 
The axis of abscissa $k_l$ is gauged in the unit of $1/d_l$ 
with $l = x, y$. 
(a) $E^{(+)}(\boldsymbol{k})$ in the $k_x-k_y$ plane for the up-spin
bands. 
(b) $E^{(-)}(\boldsymbol{k})$ in the $k_x-k_y$ plane for the down-spin
bands.
[This figure is quoted from Fig.~7 of Ref.~\onlinecite{Zhang2022} after some processing.]
}
\label{fig2}
\end{figure}

Figures~\ref{fig2}(a) and \ref{fig2}(b) show bulk energy dispersions of $E(\boldsymbol{k})$ of up-spin and down-spin FWSM bands of $e(-1)$ and $hh(+1)$ at $k_z = 0$, respectively, where $F_c=0.0003$ (1.54 MV/cm), and $\omega=$0.0147 (0.4 eV) that is almost resonant with the bandgap $E_g$.
The opening of Floquet bandgap is confirmed in both panels; in fact, these bands remain open besides $k_z=0$ as far as $-k_z^{W(\pm)} < k_z < k_z^{W(\pm)}$.
It is seen that the up-spin Floquet band of $e(-1)$ looks almost degenerate with that of $hh(-1)$, accompanying the formation of a nodal ring reminiscent of the Floquet nodal-line SM, in contrast with the down-spin Floquet band showing the clear bandgap formation.
The difference is attributed to the fact that optical transitions between bands with down-spin are more dominantly made than those with up-spin by the left-hand circularly polarized laser, as mentioned in Sec.~\ref{sec1}.
To be more specific, the band gap of the up-spin Floquet band arises from an optical
spin-orbit coupling, namely, a spin-orbit coupling $t_{sp}$ reduced by a factor of $J_1(z_c)$ with $z_c=F_cd_c/\omega$ and $d_c\equiv d_x=d_y$, while that of the down-spin band is governed by an optical dipole coupling $\Omega_c(\equiv \Omega_x=\Omega_y)$, where $t_{sp}J_1(z_c)=2.1\times 10^{-4}$ and $\Omega_c=5.5\times 10^{-3}$.\cite{Zhang2022}
As $\omega$ increases from $E_g$, the dominance of the down-spin bands becomes more suppressed, and eventually, the difference between these two would disappear for $\omega \gg E_g$.
Here is one more comment on the nodal ring of the up-spin  Floquet band.
The energy dispersion $E^{(+)}(\boldsymbol{k})$ at any $k_z$  has the similar structure as that of  Fig.~\ref{fig2}(a) at $k_z=0$ as long as $|k_z|<k_z^{W(+)}$, apart from the fact that the positions of the rings shift upward in energy from $E_F=0$ and the diameters of the rings decrease as $k_z$ increases.
Eventually, the ring converges to the Weyl node with  vanishing of the diameter. 
By connecting these rings as functions of $k_z$ continuously, one can images that the intersection between $e(-1)$ and $hh(+1)$ bands forms a manifold similar to a spheroid in the BZ with the $k_x$ and $k_y$ axes as two principle axes with the same length, and a line connecting a pair of Weyl nodes, namely, the $k_z$ axis as the third axis.

\begin{figure}[t]
\begin{center}
\includegraphics[width=8.5cm]{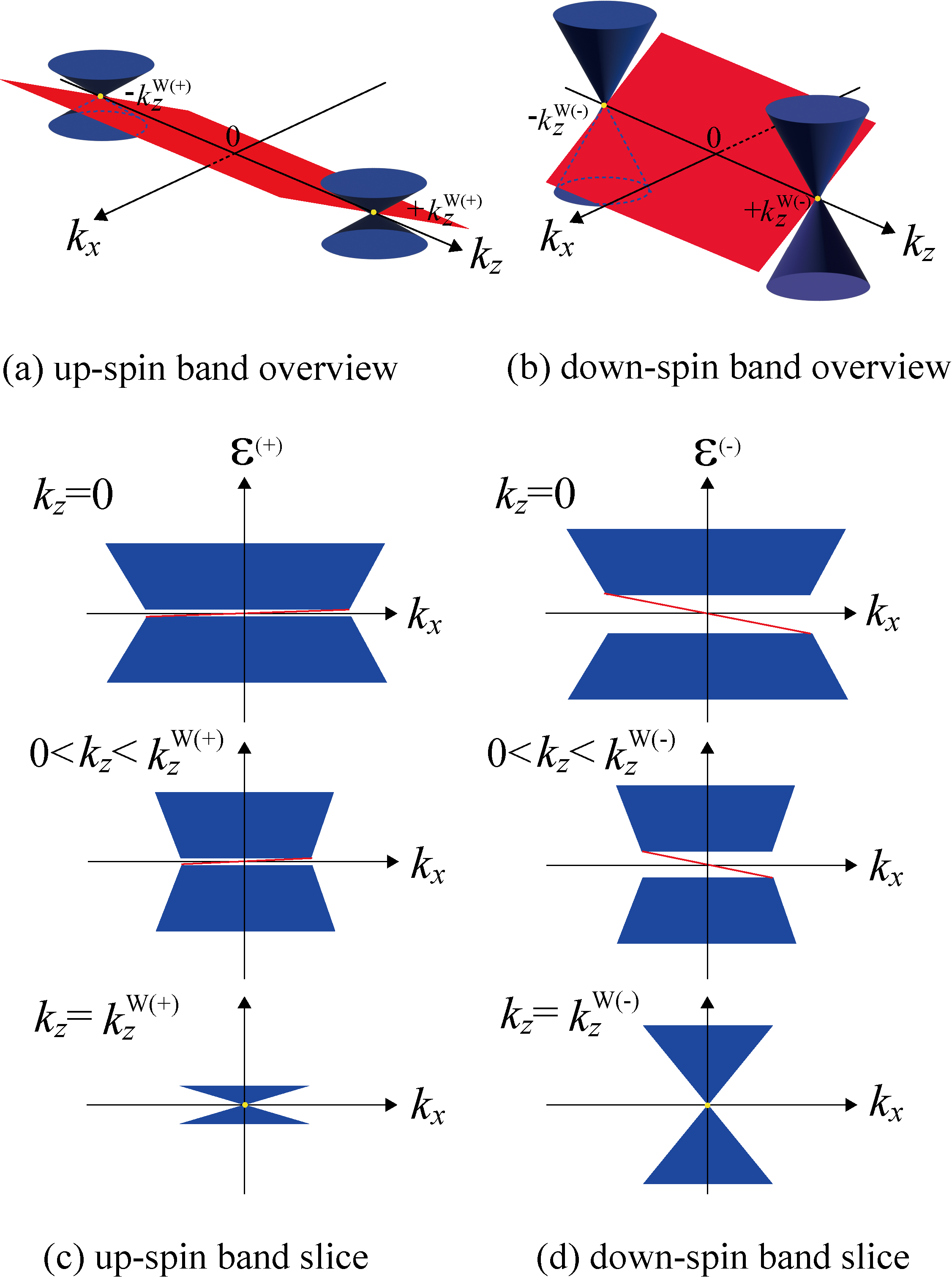}
\end{center}
\caption{Schematic view of band strutures of FWSM for both spin states.
(a) Overview of the up-spin band with a surface state (painted in red) pinned to a pair of Weyl nodes at $\pm k_z^{(W(+)}$  (specified in yellow), which are formed as intersections of Dirac cones  (painted in blue).
(b) The same as panel (a) but for the down-spin  band.
(c) Energy dispersions of the up-spin band $\mathcal{E}^{(+)}$ as functions of $k_x$, sliced at $k_z=0, k_z$ with $0 < k_z < k_z^{W(+)}$, and $k_z=k_z^{W(+)}$.
The projected bulk bands are painted in blue, and the surfaces states are depicted by red solid lines.
(d) The same as panel (c) but energy dispersions of the down-spin band $\mathcal{E}^{(-)}$.
}
\label{fig0}
\end{figure}

Hereafter, it is understood that an open boundary condition is imposed on the Floquet eigenvalue equation of Eq.~(\ref{varphi1}) in the $y$ direction in place of the periodic one, where an electron is confined within a finite ragne of $y$, while moves freely in the $x-z$ plane.
The resulting band of $\alpha$ state is represented as $\mathcal{E}_\alpha(\bar{\boldsymbol{k}})$ with $\bar{\boldsymbol{k}}=(k_x,k_z)$ to make distinct from $E_\alpha(\boldsymbol{k})$, where $\mathcal{E}_\alpha(\bar{\boldsymbol{k}})$ is considered as a projection of $E_\alpha(\boldsymbol{k})$ onto the $k_x-k_z$ plane.
The confinement of electron results in the formation of surface states within the band gaps seen in Figs.~\ref{fig2}(a) and \ref{fig2}(b).
Figure~\ref{fig0} is schematic view of FWSM band structures, $\mathcal{E}^{(+)}(\bar{\boldsymbol{k}})$ and $\mathcal{E}^{(-)}(\bar{\boldsymbol{k}})$, for the up-spin and down-spin states, respectively, which is depicted based on the results of the numerical calculations of Eq.~(\ref{varphi1}) under the open boundary condition.
Figures~\ref{fig0}(a) and \ref{fig0}(b) show the surface states  pinned to a pair of Weyl nodes at $\bar{\boldsymbol{k}}^{W(\pm)}\equiv (0,k_z^{W(\pm)})$ and $-\bar{\boldsymbol{k}}^{W(\pm)}$, and exhibit a Fermi arc if all electrons are assumed to be occupied below $E_F$.
Here it is also shown that the height of the up-spin Weyl cones is much smaller than that of the down-spin Weyl cones due to the same reason as the difference of bandgap between Figs.~\ref{fig2}(a) and \ref{fig2}(b).
In Fig.~\ref{fig0}(c)/\ref{fig0}(d), the energy dispersion $\mathcal{E}^{(+)}/\mathcal{E}^{(-)}$ of the up-spin/down-spin band is sliced at three different $k_z$'s, namely, $k_z=0$, $k_z$ located between 0 and $k_z^{W(+)}/k_z^{W(-)}$, and $k_z=k_z^{W(+)}/k_z^{W(-)}$, and these sliced energy bands are shown as functions of $k_x$.
The range of line segment in the $k_x$ axis that supports a surface state is maximized at $k_z=0$, and more reduced  with increasing  $k_z$ from 0 to $k_z^{W(\pm)}$.
Incidentally, connecting these line segments in Fig.~\ref{fig0}(c) continuously from $-k_z^{W(+)}$ to $+k_z^{W(+)}$ leads  to  a manifold similar to an ellipse formed in the $k_x-k_z$ plane; obviously, this is nothing but the projection of the manifold mentioned above for $E^{(+)}(\boldsymbol{k})$.

These surface states play decisive roles of generating solenoid magnetization $\bar{\boldsymbol{M}}^{(sol)}(t)$ of Eq.~(\ref{Msol2}) in this study, as shown below.
This is because the main part of the spin magnetization $\bar{\boldsymbol{M}}(t)$ is caused due to the imbalance between excited carrier densities in the up-spin band and those in the down-spin band.
Here, the degree of imbalance corresponds to the degree of spin polarization, denoted as $I_{spin}$, which is defined as the ratio of the spin difference to the total number of electrons.
Such imbalance is considered more conspicuous than that in the bulk states, since most of electrons in the bulk still stay at the parent band $hh(0)$ associated with the ground state in spite of the intense optical excitation of concern.

\subsection{Surface magnetization}
\label{sec3B}

The excited electron densities at time $t$ in the bulk and surface are evaluated by the volume average density $\eta_{ee}^{(b)}(t)$ and the surface average density $\eta_{ee}^{(s)}(t)$, respectively.
These are the diagonal components of $\eta_{\mu\nu}^{(b)}(t)$ and $\eta_{\mu\nu}^{(s)}(t)$ defined in terms of $\rho_{\boldsymbol{k}\mu\nu}(t)$, as shown in Eqs.~(\ref{etabC}) and (\ref{etasC}), respectively.
The surface magnetization is provided by replacing Eqs.~(\ref{Mbar}) and (\ref{Mbar2}) by alternative expressions suitable for the generation of the surface states, which is read as Eq.~(\ref{MsolsurfC}), that is, 
\begin{equation}
\bar{\boldsymbol{M}}^{(sol)}(t)
=
\mu_B\Re
\sum_{\mu\nu}
\eta^{(s)}_{\mu\nu}(t)
\langle\mu|g_e\boldsymbol{s}+\boldsymbol{l}+\boldsymbol{r}
\times\boldsymbol{A}(t)|\nu\rangle.
\label{Msolsurf}
\end{equation}
It is found that this is composed of the three contributions from the spin magnetization, the orbital magnetization relevant to the IFE, and the remaining term due to the gauge correction.
Here, the $z$ component of it, $\bar{M}_z^{(sol)}(t)$, is evaluated.

\begin{figure}[t]
\begin{center}
\includegraphics[width=8.5cm]{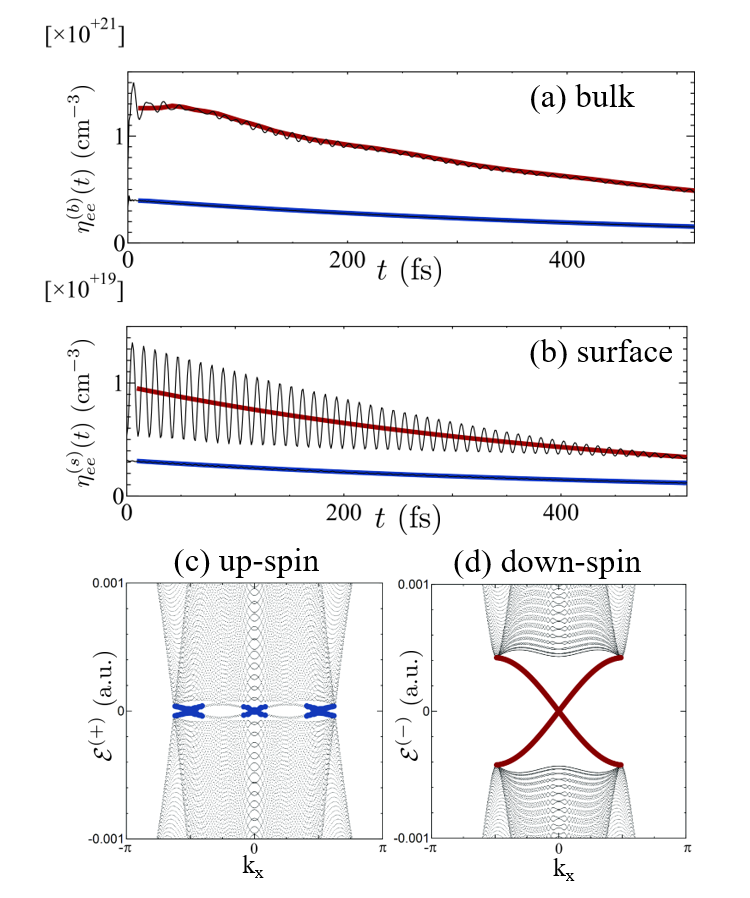}
\end{center}
\caption{
(a) The excited electron densities in the bulk $\eta_{ee}^{(b)}(t)$ (in the unit of cm$^{-3}$) for both spins as functions of time $t$ (in the unit of fs), where $F_c=0.0003$ (1.54 MV/cm) and $\omega=$0.0147 (0.4 eV).
The red and blue solid lines represent one-cycle time average of these electron densities of down spin and up-spin, respectively.
(b) The excited electron densities in the surface $\eta_{ee}^{(s)}(t)$ (cm$^{-3}$) for both spins as functions of $t$ (fs), where $F_c=0.0003$ (1.54 MV/cm) and $\omega=$0.0147 (0.4 eV). 
Here, these densities are gauge in the unit of cm$^{-3}$ rather than cm$^{-2}$.
For more detail, consult the text.
(c) Surface states incorporated with the calculation of $\eta_{ee}^{(s)}(t)$ for the up spin, represented by blue solid lines, are shown overlaid on $\mathcal{E}^{(+)}(\bar{\boldsymbol{k}})$ as a function of $k_x$ with $k_z=0$.
(d) The same as panel (c) but for the down spin, represented by red solid lines, are shown overlaid on $\mathcal{E}^{(-)}(\bar{\boldsymbol{k}})$ as a function of $k_x$ with $k_z=0$.
}
\label{fig3}
\end{figure}

Figure~\ref{fig3}(a) shows $\eta_{ee}^{(b)}(t)$'s for both spins as functions of $t$ with the relaxation times introduced in Eqs.~(\ref{OBED}) and (\ref{OBEoffD}) as $T_1=500$ fs and $T_2=250$ fs, respectively, where $F_c=0.0003$ (1.54 MV/cm) and $\omega=$0.0147 (0.4 eV).
The total valence electron density of the crystal amounts to $n_e=4.4 \times 10^{22}$ cm$^{-3}$.
The red and blue solid lines represent one-cycle time $(T)$ average of these electron densities of down spin and up-spin, respectively, and $I_{spin} \approx 0.01$ at $t=25T$; where one period equals $T=$ 10.3 fs.
Figure~\ref{fig3}(b) shows $\eta_{ee}^{(s)}(t)$'s for both spins as functions of $t$, where surface electron states are extracted from all of the calculated Floquet states numerically by inspecting the degree of electron densities localized at the outermost site in the $y$ direction.
The incorporated surface states are shown in Fig.~\ref{fig3}(c) and \ref{fig3}(d).
It is remarked that for the purpose of comparison with $\eta_{ee}^{(b)}(t)$ of Fig.~\ref{fig3}(a), $\eta_{ee}^{(s)}(t)$ of Fig.~\ref{fig3}(b) is shown in the unit of cm$^{-3}$ by dividing $\eta_{ee}^{(s)}(t)$ defined in Eq.~(\ref{etasC}) by the lattice constant $d_y=3$\AA$\:$ in the $y$ direction instead of the usual unit of cm$^{-2}$.
Here, it is found that $I_{spin} \approx 0.5$ at $t=25T$ that is much greater than that in the bulk.
The surface electron densities created by the drive with the linear polarization show no spin difference between both excited electron densities, as is clear (though not shown here).

\begin{figure}[t]
\begin{center}
\includegraphics[width=8.5cm]{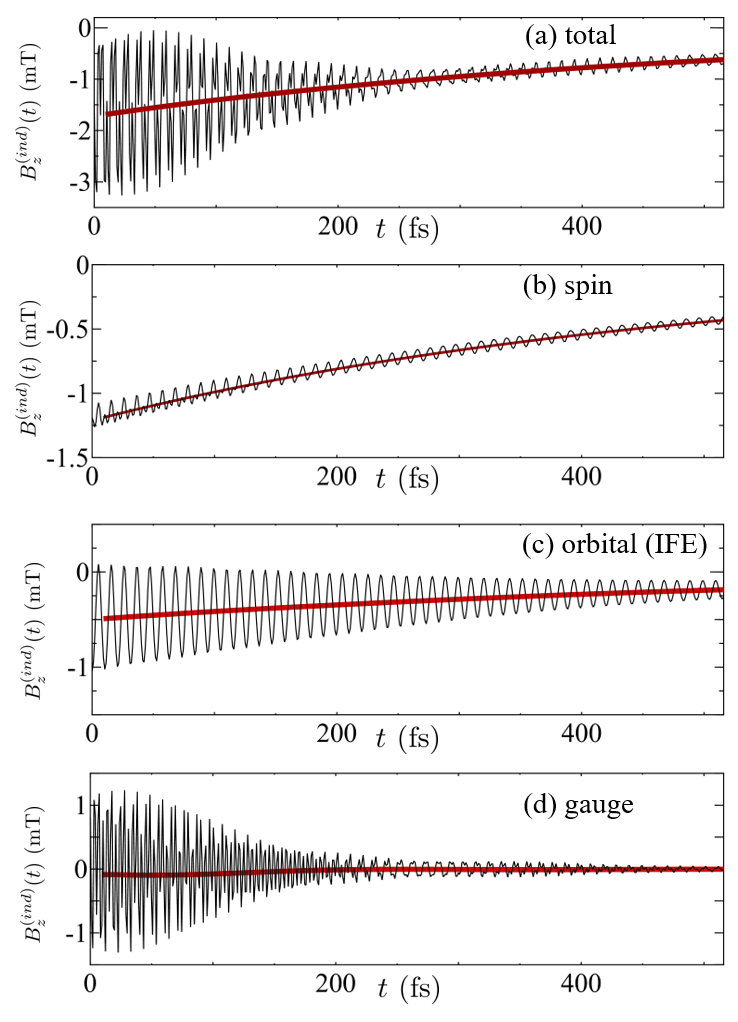}
\end{center}
\caption{
(a) The surface magnetic induction $B_z^{(ind)}(t)$ (in the unit of mT), represented by a solid black line, as a function of time $t$ (in the unit of fs), where $F_c=0.0003$ (1.54 MV/cm) and $\omega=$0.0147 (0.4 eV).
Here, the one-cycle average value is represented by a solid red line.
For more detail, consult the text.
(b) The same as panel (a), but for the surface magnetization attributed to the spin magnetization.
(c) The same as panel (a), but for the surface magnetization attributed to the orbital magnetization associated with the IFE.
(d) The same as panel (a), but for the surface magnetization attributed to the gauge correction term.
}
\label{fig4}
\end{figure}

Figure~\ref{fig4}(a) shows the calculated result of the induced surface magnetization (represented by a solid black line), where the one-cycle average value is represented by a solid red line, and the magnetic induction, namely, the magnetization multiplied by $\mu_0$, $B_z^{(ind)}(t)\equiv\mu_0\bar{M}_z^{(sol)}(t)$, is gauged in the unit of mT, where $\eta_{\mu\nu}^{(s)}(t)$ appearing in Eq.~(\ref{Msolsurf}) is redefined by dividing by $d_y$. 
It is found that the magnetization of the order of 1 mT is induced in the present system.
This strength seems compatible with the rough estimate that $\eta_{ee}^{(s)}(t)$ is about one thousand times smaller than $n_e$, where assuming all electrons of the crystal are aligned as an ideal ferromagnet, the magnetization of the order of 1 T is generated besides the IFE.
In passing, the strength of a refrigerator magnet is of the order of mT.
The detailed origins of the induced magnetization are analyzed by examining the respective contributions from the three different generation mechanisms, which are shown in Figs.~\ref{fig4}(b)-\ref{fig4}(d).
It is seen that the negative values of the spin magnetization and the orbital magnetization, namely, the IFE, are in harmony with the dominance of $\eta_{ee}^{(s)}(t)$ of the down spin with $J_z=-1/2$ over that of the up spin with $J_z=1/2$, as shown in Fig.~\ref{fig3}(b).
To be more precise, $\eta_{hh}^{(s)}(t)$ of the down spin with $J_z=-3/2$ also contributes to these magnetization, since the down-spin Floquet bands are formed by the hybridization between the two bands with $J_z=-1/2, -3/2$ through both of the spin-orbit and electric-dipole couplings.
The contributions of the IFE to $B_z^{(ind)}(t)$ are always slightly smaller than those of the spin magnetization regardless of $F_c$ to the extent examined. 
On the other hand, the magnetization due to the gauge correction oscillates violently immediately after the excitation, but asymptotically becomes negligibly small.
Such tendency of this correction term remains largely unchanged regardless of $F_c$ to the extent examined.
On the other hand, it would be worth mentioning here the induced magnetization of bulk.
According to numerical calculations of $\bar{\boldsymbol{M}}^{(sol)}(t)$ with modification from Eq.~(\ref{Msolsurf}) for the bulk, the induced magnetization is estimated to be about 40 mT at $t=25 T$ as a mere ideal maximum.\cite{comment1}
The most contribution comes from the IFE, though both of the spin magnetization and the gauge correction have little effects.

Here, the surface magnetic induction of the IFE imparted by $B_z^{(ind)}$, which is represented as $B_z^{(IFE)}$, is discussed in a qualitative manner in terms of the expression of IFE based on the Drude-Lorentz model.
This is read as\cite{Battiato2014}
\begin{equation}
B_z^{(IFE)}=-{\mu_0n_e^{(s)}(F_c)\over 2}{\omega\over (\omega^2-\omega_0^2)^2+(\gamma\omega)^2}F_c^2,
\end{equation}
where $n_e^{(s)}(F_c)$ represents a surface electron density, $\omega_0$ and $1/\gamma$ represent a resonance energy and a transverse relaxation time, respectively.
It is noted that the electrons occupied in the surface states are composed of a portion of excited carriers in the present FWSM system, and thus the leading contribution of $n_e^{(s)}(F_c)$ arises from $F_c^2$ terms, namely, $n_e^{(s)}(F_c)\propto F_c^2$.
This implies that $B_z^{(IFE)}$ of concern would be dominated by the $F_c^4$ terms, differing from the conventional IFE depending on $F_c^2$.
Here, it would be appropriate to set $\omega_0$ equal to zero, since the FWSM surface states are considered metallic.
Thus, with $n_e^{(s)}(F_c)\approx 0.5\times 10^{19}$ cm$^{-3}$, $\omega=0.4$ eV, $F_c=1.54$ MV/cm, and $1/\gamma=T_2=250$ fs, one obtains $B_z^{(IFE)}\approx$ -2.5 mT.
The results of Fig.~{\ref{fig4}(c)} are somehow consistent with this estimate.
Incidentally, the magnitude of it is about  similar to that of a disordered metal, though under a different optical condition.\cite{Taguchi2011}

\begin{figure}[t]
\begin{center}
\includegraphics[width=8.5cm]{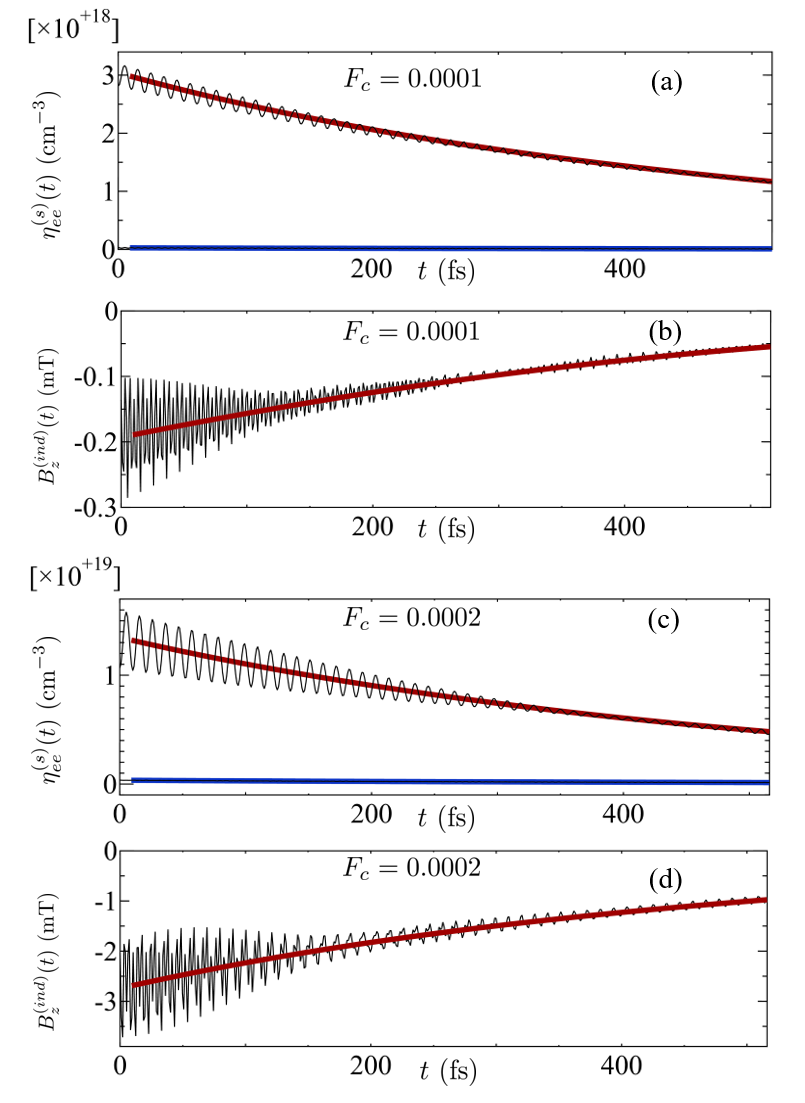}
\end{center}
\caption{
The $F_c$ dependence of the excited electron densities in the surface $\eta_{ee}^{(s)}(t)$ and the surface magnetic induction $B_z^{(ind)}(t)$ with $\omega=0.0147$ (0.4 eV).
(a) and (b) The same as Fig.~\ref{fig3}(b) and Fig.~\ref{fig4}(a), respectively, but for $F_c=0.0001$ (0.51 MV/cm).
(c) and (d) The same as Fig.~\ref{fig3}(b) and Fig.~\ref{fig4}(a), respectively, but for $F_c=0.0002$ (1.03 MV/cm).
}
\label{fig5}
\end{figure}

Next, the discussion is made on the $F_c$ dependence of the surface magnetization with $\omega=0.0147$ (0.4 eV).
The variance of locations of the Weyl nodes for both spin states with respect to $F_c$ is shown in Fig.~\ref{fig1}(a), where it is seen that the Weyl node moves from the band center at $k_z=0$ toward the band edge as $F_c$ decreases.
Figures~\ref{fig5}(a) and \ref{fig5}(b) show $\eta_{ee}^{(s)}(t)$ and $B_z^{(ind)}(t)$ at $F_c=0.0001$ (0.51 MV/cm), respectively, and Figs.~\ref{fig5}(c) and \ref{fig5}(d) show those at $F_c=0.0002$ (1.03 MV/cm).
These correspond to Figs.~\ref{fig3}(b) and \ref{fig4}(a) for $F_c=0.0003$.
The projected Floquet bands $\mathcal{E}(\bar{\boldsymbol{k}})$ in both cases of $F_c$ look similar to those of Figs.~\ref{fig3}(c) and \ref{fig3}(d) besides the size of band gap at $k_z=0$ (though not shown here).
It is obvious that $\eta_{ee}^{(s)}(t)$ at $F_c=0.0002$ is much dominant over that at $F_c=0.0001$, which is reflected on the difference between $B_z^{(ind)}(t)$'s in both cases in a straightforward manner.
It is noted that the magnitude of $B_z^{(ind)}(t)$ at $F_c=0.0002$ is noticeably larger than that at $F_c=0.0003$.
This is understood from the perspective of the degree of spin polarization $I_{spin}$, that is, $I_{spin}\approx 1$ at $F_c=0.0002$ even though $\eta_{ee}^{(s)}(t)$ for the down spin is of the comparable order with that at $F_c=0.0003$.
As $F_c$ increases from 0.0003, the magnitude of $B_z^{(ind)}(t)$ becomes smaller.
Further, the direction of it changes to the positive sign when $F_c$ exceeds around 0.0004, since the pair of Weyl nodes of the down spin is annihilated at $k_z=0$ to cause the disappearance of the surface states, as shown in Fig.~\ref{fig1}(a).

\begin{figure}[t]
\begin{center}
\includegraphics[width=8.5cm]{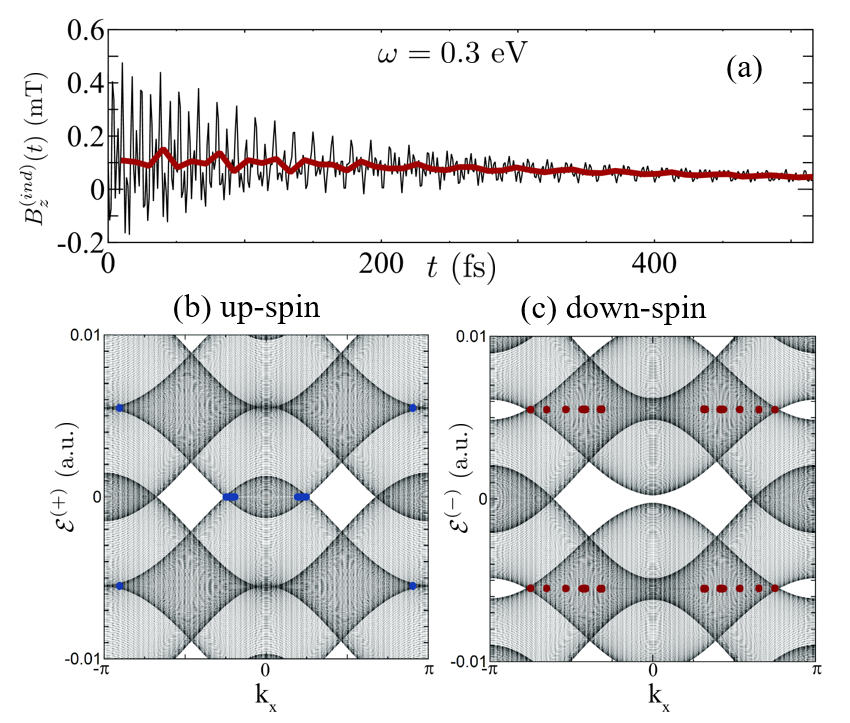}
\end{center}
\caption{
The $\omega$ dependence of the surface magnetic induction $B_z^{(ind)}(t)$ with $F_c=0.0003$ (1.54 MV/cm).
(a) The same as Fig.~\ref{fig4}(a) but for $\omega=0.0110$ (0.3 eV).
(b) Surface states incorporated with the calculation of $B_z^{(ind)}(t)$ for the up spin, represented by blue dots, similar to Fig.~\ref{fig3}(c).
(c) Surface states incorporated with the calculation of $B_z^{(ind)}(t)$ for the down spin, represented by red dots, similar to Fig.~\ref{fig3}(d).
}
\label{fig6}
\end{figure}

As regards the $\omega$ dependence of the surface magnetization with $F_c=0.0003$ (1.54 MV/cm), the variance of locations of the Weyl nodes for both spin states with respect to $\omega$ is shown in Fig.~\ref{fig1}(b), where it is seen that the Weyl node moves from the band center toward the band edge as $\omega$ increases.
Figure~\ref{fig6}(a) shows $B_z^{(ind)}(t)$ at $\omega=0.0110$ (0.3 eV).
Here, the up spin band forms a pair of Weyl nodes, whereas a gap still opens in the down spin band, accompanying a trivial insulating phase, as seen in Figs.~\ref{fig6}(b) and \ref{fig6}(c).
This is consistent with the observation that the resulting magnetization is oriented in a positive direction with much smaller magnitude than that seen in Fig.~\ref{fig4}(a) at $\omega=0.4$ eV.

\begin{figure}[t]
\begin{center}
\includegraphics[width=8.5cm]{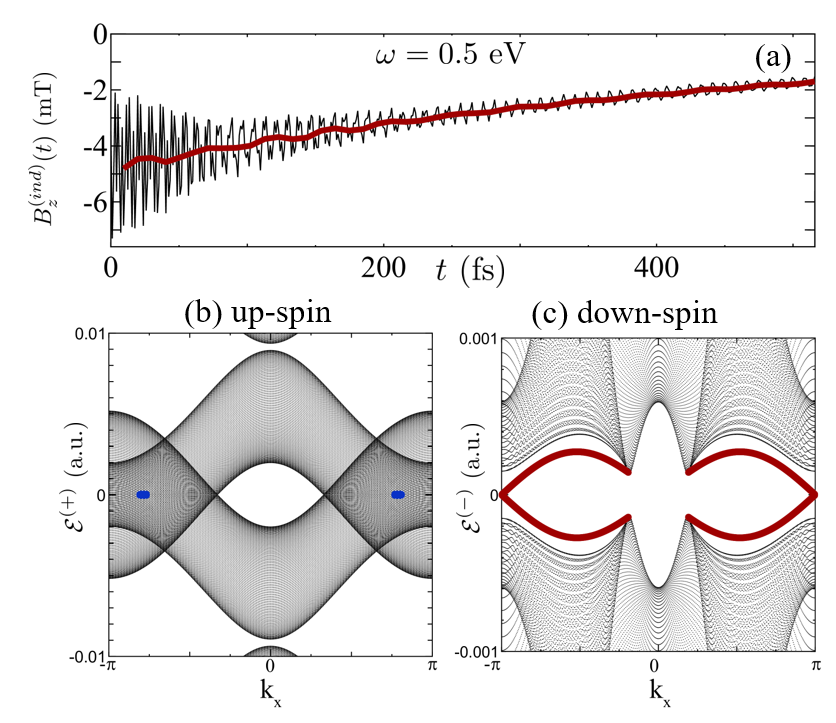}
\end{center}
\caption{
The $\omega$ dependence of the surface magnetic induction $B_z^{(ind)}(t)$ with $F_c=0.0003$ (1.54 MV/cm).
(a) The same as Fig.~\ref{fig6}(a) but for $\omega=0.0184$ (0.5 eV).
(b) The same as Fig.~\ref{fig6}(b) but for $\omega=0.0184$ (0.5 eV).
(c) The same as Fig.~\ref{fig6}(c) but for $\omega=0.0184$ (0.5 eV).
}
\label{fig7}
\end{figure}

As $\omega$ increases above the band gap $E_g=0.46$ eV, the picture changes dramatically, as shown for $\omega=0.5$ eV in Fig.~\ref{fig7}.
As mentioned in Sec.~\ref{sec3A}, the Floquet bands of $\mathcal{E}_{e(-1)}(\bar{\boldsymbol{k}})$ and $\mathcal{E}_{hh(+1)}(\bar{\boldsymbol{k}})$ form the pairs of Weyl nodes at $k_z=k_z^{(\pm)}, -k_z^{(\pm)}$ accompanying the surface states for $\omega <E_g$.
Here, bands are arranged in ascending order in energy as $hh(0),e(-1),hh(+1)$, and $e(0)$ in the whole two-dimensional BZ.
However, when $\omega > E_g$, it is expected that these bands are rearranged in the order of $e(-1),hh(0),e(0)$, and $hh(+1)$ in the limited region of the BZ including $\bar{\boldsymbol{k}}=\boldsymbol{0}$.
Figure~\ref{fig7}(c) for the down-spin bands is the case.
Here, a pair of bands of $hh(+1)$ and $e(0)$ repel with each other strongly, forming a conspicuous anti-crossing due to the optical dipole coupling without surface states, so do another pair of bands of $hh(0)$ and $e(-1)$.
Further, the bottom of band $e(0)$ lies above the top of band $h(0)$, forming a band gap around $E_F=0$.
In the remaining region of the BZ, a band gap is still formed by the bands of $e(-1)$ and $hh(+1)$ around $E_F=0$, supporting surface states extending to the edges of the BK.
As regards the up-spin bands seen in Fig.~\ref{fig7}(b), the band rearrangement does not occur yet, and the bottom of band $hh(+1)$ lies above the top of band $e(-1)$, which reduces the region of the BK where the surface states are allowed to exist.
The resulting magnetization is shown in Fig.~\ref{fig7}(a), and this is larger than that in Fig.~\ref{fig5}(d), presumably due to enhancement of $I_{spin}$.

Roughly speaking, the magnetization $B^{(ind)}_z(t)$ is considered proportional to the product of $\eta^{(s)}_{ee}(t)$ of down-spin electrons and an area of the associated surface state $A_W$, where $A_W$ corresponds to the area in the BZ that is painted in red in Figs.~\ref{fig0}(b).
Assuming that the shape of the surface state in the BZ is an ellipse, as mentioned in Sec.~\ref{sec3A}, this area is approximately represented as 
\begin{equation}
A_W={\pi K_xK_z\over \cos\theta_W},
\end{equation}
where $K_z$ is half of the separation of the pair of Weyl nodes, given by $K_z=k_z^{W(-)}$, $K_x$ is half of the line segment in the $k_x$ axis that supports the down-spin surface state at $k_z = 0$, and $\theta_W$ is an intersection angle between  the down-spin surface plane in the BZ and the $k_x-k_z$ plane.
That is, 
\begin{equation}
B^{(ind)}_z(t) \propto \eta^{(s)}_{ee}(t)A_W.
\label{rough}
\end{equation}
This is considered correct under restricted conditions that $\omega \approx E_g$ and $F_c \lesssim 0.0003$, corresponding to the cases of Figs.~\ref{fig4}(a), \ref{fig5}(a), and \ref{fig5}(c).
Otherwise,  the expression of $A_W$ can be modified to a certain extent, and $B^{(ind)}_z(t)$ would be of more complicated form.
Actually, anti-crossings formed in the region of $\omega > E_g$ hinders the formation of surface state around $k_x=0$ in part, which corresponds to the case of Fig.~\ref{fig7}(a).
However, the separation $K_z$ would still plays an important role of the formation of magnetization, since $K_z$ becomes greater with the increase of $\omega$, as seen in Fig.~\ref{fig1}(b).
In summary, it is speculated that $K_z$ is a crucial factor governing the surface magnetization.
Incidentally, this factor also plays a key role of the anomalous Hall effect that is expected to be realized in the present FWSMs,\cite{Wang2013,Gynther2011,Son2013,Arnold2016,Huang2015,Ali2014}
 though this is beyond the scope of this paper.

\section{Conclusions}
\label{sec4}

The nonmagnetic and narrow-gap semiconductor Zn$_3$As$_2$ is driven by a left-hand circularly-polarized continuous-wave laser with a nearly resonant frequency to the band gap to produce the two types of FWSM phases distinguished by the spins.
The numerical evaluation of the transient non-equilibrium dynamics of the concerned system shows that the FWSM surface state formation accompanies the surface magnetization composed of the spin polarization and the orbital magnetization corresponding to the IFE with the gauge correction.
The magnitude and direction of the magnetization varies sharply, depending on the intensity and frequency of the driving laser.
It is stressed that the surface magnetization found here is more pronounced, when the laser has a nearly resonant frequency, and this is strong enough to form distinct Weyl pairs in the BK and to make the occupied electrons in the surface states highly spin-polarized.
Such an intriguing phenomenon would be measurable by virtue of the longitudinal magneto-optic Kerr effect appropriate for detecting surface magnetization.\cite{Kerr1877,Weinberg2008,Haider2017}
However, it is remarked that in actual experiments, high-intensity laser employed here would possibly lead to effects of disturbing the predicted magnetization due to heating, spin relaxation through a coupling with phonons and a spin-spin interaction, band gap renormalization, and so on, which are not considered here.

The findings in this paper have the following novelty and significance.
The laser-induced magnetization presented here is a departure from conventional research on photo-induced magnetization \cite{Kimel2005,Vahaplar2009,Ostler2012,Kurkin2008,Taguchi2011,Popova2011,Popova2012,Gridnev2013,Qaiumzadeh2013,Yoshikawa2022} in that in the former, the surface magnetization separated from bulk magnetization is enhanced by spin-magnetization due to high-density resonant excitation in addition with the IFE, while in the latter, this is  due to spin-flips simply caused by  the IFE in coordination with  a strong spin-orbit coupling.
Such induced magnetization is expected to provide a new platform for the studies of magnetotransport in topological materials.
For instance, this would enable optical control of chiral anomaly that is still a less studied issue.\cite{MehdiJadidi2020}
To be more specific, it is likely that the surface magnetization is further fed back to an interaction with electronic states of a pair of FWSMs to form Landau levels in the $k_z$-direction in a realistic situation; though this is disregarded in the theoretical model employed here.
Application of an additional electric field along the $z$-direction is considered to exhibit negative magnetic resistance due to chiral anomaly that is dominantly shown in a relatively low magnetic field region corresponding to that evaluated here. \cite{Son2013,Burkov2014,Burkov2015,Xiong2015,Zhang2016a}

Finally, one more comment is made on the advantage of the theoretical method developed here to deal with the non-equilibrium dynamics of the Floquet topological systems.
To deepen the understanding of this systems in non-equilibrium and to further explore the underlying physics of it, an approach from the transient quantum dynamics accompanying real carrier excitation is inevitably required in addition with just the analyses of Floquet band structures.
In this study, by introducing the creation and annihilation operators of a fermionic quasiparticle, represented as $\hat{b}^\dagger_{\boldsymbol{k}\alpha}(t)$ and $\hat{b}_{\boldsymbol{k}\alpha}(t)$, respectively, a density matrix $\rho_{\boldsymbol{k}\mu\nu}(t)=\langle \hat{a}^\dagger_{\boldsymbol{k}\mu}(t)\hat{a}_{\boldsymbol{k}\nu}(t) \rangle_g$ can be represented in a closed form without numerically solving the optical Bloch equations.
Here, the fermionic quasiparticle is considered as a composite particle in which laser-electron interactions are fully renormalized to a band electron in terms of the Floquet base.
Thus, it is expected that these fermionic quasiparticle operators can be employed to facilitate the description of the non-equilibrium dynamics of the topological materials driven by a laser pulse with a relatively long temporal width of the order of ps.
Actually, a variety of the Keldysh Green functions\cite{Schafer2002} are described in closed forms in a non-perturbative manner with respect to a laser-electron interaction, and these are specifically applied to high-harmonic generation, high density excitation, etc.
Further, based on these operators, it is also possible to construct the Green functions incorporating electron correlation effects for examining many-body effects of Floquet topological materials that still remain unexplored.

\begin{acknowledgments}
This work was supported by JSPS KAKENHI Grant No. JP19K03695, and the University of Tsukuba, Pre-Strategic Initiatives `Development Center for High-Function and High-Performance Organic-Inorganic Spin Electronics'.
\end{acknowledgments}

\appendix
\section{Miscellaneous relations derived from the Floquet eigenvalue equations}
\label{app1}

The Floquet eigenvalue equations are represented by the following equations as
\begin{equation}
\sum_{\nu}
L_{F\mu\nu}(\boldsymbol{k},t)\varphi_{\nu,\boldsymbol{k}\alpha}(t)=E_\alpha(\boldsymbol{k})\varphi_{\mu,\boldsymbol{k}\alpha}(t),
\label{Avarphi1}
\end{equation}
and
\begin{equation}
\sum_{\nu n^\prime}L_{F\:\mu n,\nu n^\prime}(\boldsymbol{k})C_{\nu n^\prime,\boldsymbol{k}\alpha}
=E_\alpha(\boldsymbol{k}) C_{\mu n,\boldsymbol{k}\alpha},
\label{AEqC}
\end{equation}
given by Eqs.~(\ref{varphi1}) and (\ref{EqC}), respectively, where
\begin{equation}
\varphi_{\mu,\boldsymbol{k}\alpha}(t)=\sum_ne^{in\omega t}C_{\mu n,\boldsymbol{k}\alpha}.
\label{Avarphi2}
\end{equation}
There are the unitarity relations regarding the matrix $\{C_{\mu n, \alpha^\prime}\}$,
given by
\begin{equation}
\sum_{\mu n} C^*_{\mu n,\boldsymbol{k}\alpha}C_{\mu n, \boldsymbol{k}\alpha^\prime}
=\delta_{\alpha\alpha^\prime}
\label{R1}
\end{equation}
and
\begin{equation}
\sum_\alpha C_{\nu n,\boldsymbol{k}\alpha}C^*_{\mu n^\prime, \boldsymbol{k}\alpha}
=\delta_{\mu\nu}\delta_{nn^\prime}.
\label{R2}
\end{equation}
Here, the orthonormalization relation for $\varphi_{\nu,\boldsymbol{k}\alpha}(t)$ is given by
\begin{equation}
{1\over T}\int^T_0 dt \sum_\mu \varphi^*_{\mu,\boldsymbol{k}\alpha}(t)\varphi_{\mu,\boldsymbol{k}\alpha^\prime}(t)=\delta_{\alpha\alpha^\prime}.
\end{equation}
and the associated closure relation is given by
\begin{equation}
{1\over N_{ph}}\sum_\alpha \varphi_{\nu,\boldsymbol{k}\alpha}(t)\varphi^*_{\mu,\boldsymbol{k}\alpha}(t)=\delta_{\nu\mu},
\label{R4}
\end{equation}
where
$N_{ph}=\sum_{N_\alpha} 1$.
This is readily verified as follows:
\begin{eqnarray}
&&{1\over N_{ph}}\sum_\alpha \varphi_{\nu,\boldsymbol{k}\alpha}(t)\varphi^*_{\mu,\boldsymbol{k}\alpha}(t)
\nonumber\\
&=&{1\over N_{ph}}\sum_{\alpha,n,n^\prime}e^{i(n-n^\prime)\omega t}
C_{\nu n,\boldsymbol{k}\alpha}C^*_{\mu n^\prime,\boldsymbol{k}\alpha}
\nonumber\\
&=&
{1\over N_{ph}}\sum_{n,N}e^{iN\omega t}
\delta_{\nu\mu}\delta_{N0}
\nonumber\\
&=&
\delta_{\nu\mu}{1\over N_{ph}}\sum_n 1=\delta_{\nu\mu}.
\label{R3}
\end{eqnarray}

Further, it is shown that Eq.~(\ref{Avarphi1}) becomes
\begin{eqnarray}
&&\sum_{\nu}
L_{F \mu\nu}(\boldsymbol{k},t)\left(e^{iN\omega t} \varphi_{\nu,\boldsymbol{k}\alpha}(t)\right)
\nonumber\\
&=&(E_{\alpha}(\boldsymbol{k})+N\omega)\left(e^{iN\omega t}\varphi_{\mu,\boldsymbol{k}\alpha}(t)\right)
\nonumber\\
&=&
E_{\alpha(N)}(\boldsymbol{k})\left(e^{iN\omega t}\varphi_{\mu,\boldsymbol{k}\alpha}(t)\right)
\end{eqnarray}
with
$
E_{\alpha(N)}(\boldsymbol{k})\equiv E_{\alpha}(\boldsymbol{k})+N\omega,
$
and thus one obtains the relation that
\begin{equation}
e^{iN\omega t} \varphi_{\nu,\boldsymbol{k}\alpha}(t)=\varphi_{\nu,\boldsymbol{k}\alpha(N)}(t).
\label{Avarphi3}
\end{equation}
Here, it is understood that the photon sideband $\alpha$ is represented as
$\alpha_0(N_\alpha)$, where $\alpha_0(\equiv\alpha_0(0))$ and $N_\alpha$ stand for a parent band 
and the number of the sideband, respectively.
Further, 
$\alpha(N)\equiv\alpha_0(N_\alpha+N)$
and 
$E_{\alpha(N)}(\boldsymbol{k})=E_{\alpha_0}(\boldsymbol{k})+(N_\alpha+N)\omega$.
In view of Eq.~(\ref{Avarphi2}), Eq.~(\ref{Avarphi3}) is recast into
\begin{eqnarray}
\varphi_{\nu,\boldsymbol{k}\alpha}(t)
&=&\sum_ne^{in\omega t}C_{\nu n,\boldsymbol{k}\alpha}
\nonumber\\
&=&
e^{-iN\omega t} \varphi_{\nu,\boldsymbol{k}\alpha(N)}(t)
=\sum_{n}e^{i(n-N)\omega t}C_{\nu n,\boldsymbol{k}\alpha(N)}
\nonumber\\
&=&
\sum_{n^\prime}e^{in^\prime\omega t}C_{\nu (n^\prime+N),\boldsymbol{k}\alpha(N)},
\end{eqnarray}
leading to the relation that
\begin{equation}
C_{\nu n,\boldsymbol{k}\alpha}=C_{\nu (n+N),\boldsymbol{k}\alpha(N)}.
\label{AC1}
\end{equation}

According to Eq.~(\ref{bahat}), given by
\begin{equation}
\hat{b}_{\boldsymbol{k}\alpha}^\dagger(t)=\sum_\mu\hat{a}_{\boldsymbol{k}\mu}^\dagger(t)\varphi_{\mu,\boldsymbol{k}\alpha}(t)\theta(t),
\label{Abahat}
\end{equation}
the operator $\hat{b}_{\boldsymbol{k}\alpha}^\dagger(t)$ pertinent to the creation of Floquet quasiparticle is subject to the relation
\begin{equation}
e^{iN\omega t} \hat{b}^\dagger_{\boldsymbol{k}\alpha}(t)=\hat{b}^\dagger_{\boldsymbol{k}\alpha(N)}(t)
\label{Ab1}
\end{equation}
due to Eq.~(\ref{Avarphi3}).
By employing Eqs.~(\ref{Avarphi2}) and (\ref{R2}),
Eq.~(\ref{Abahat}) is rewritten as
\begin{eqnarray}
&&\sum_\alpha \hat{b}^\dagger_{\boldsymbol{k}\alpha}(t)e^{-in\omega t}C^*_{\nu n,\boldsymbol{k}\alpha}
\nonumber\\
&=&
\sum_{\alpha\mu} \hat{a}^\dagger_{\boldsymbol{k}\mu}(t)e^{-in\omega t}\varphi_{\mu,\boldsymbol{k}\alpha}(t)
C^*_{\nu n,\boldsymbol{k}\alpha}\theta(t)
\nonumber\\
&=&
\sum_{\alpha\mu n^\prime} \hat{a}^\dagger_{\boldsymbol{k}\mu}(t)e^{i(-n+n^\prime)\omega t}
C_{\mu n^\prime,\boldsymbol{k}\alpha}C^*_{\nu n,\boldsymbol{k}\alpha}\theta(t)
\nonumber\\
&=&
\hat{a}^\dagger_{\boldsymbol{k}\nu}(t).
\end{eqnarray}
Hence, in view of Eqs.~(\ref{Ab1}) and (\ref{AC1}),
the inverse-transformed relation from Eq.~(\ref{Abahat}) is obtained as
\begin{eqnarray}
\hat{a}^\dagger_{\boldsymbol{k}\nu}(t)
&=&\sum_{\alpha(-n)} \hat{b}^\dagger_{\boldsymbol{k}\alpha(-n)}(t)C^*_{\nu 0,\boldsymbol{k}\alpha(-n)}
\nonumber\\
&=&
\sum_{\alpha} \hat{b}^\dagger_{\boldsymbol{k}\alpha}(t)C^*_{\nu 0,\boldsymbol{k}\alpha}.
\label{Arhoa2rhob}
\end{eqnarray}

Here, the matrix element $H_{F \lambda\mu}(\boldsymbol{k},t)$ is rewritten in terms of the Floquet eigenstates.
One begins with Eq.~(\ref{Avarphi1}) of the form
\begin{eqnarray}
&&\sum_\nu H_{F \lambda\nu}(\boldsymbol{k},t)\varphi_{\nu\,\boldsymbol{k}\alpha}(t)
\nonumber\\
&=&
\varphi_{\lambda,\boldsymbol{k}\alpha}(t)E_\alpha(\boldsymbol{k})
+i{d\varphi_{\lambda,\boldsymbol{k}\alpha}(t)\over dt}.
\label{varphix}
\end{eqnarray}
Multiplying the left-hand side of this by $\varphi^*_{\mu,\boldsymbol{k}\alpha}(t)$, followed by taking the summation over $\alpha$ leads to
\begin{equation}
{1\over N_{ph}}\sum_{\nu\alpha}
H_{F\lambda\nu}(\boldsymbol{k},t)\varphi_{\nu,\boldsymbol{k}\alpha}(t)\varphi^*_{\mu,\boldsymbol{k}\alpha}(t)=H_{F\lambda\mu}(\boldsymbol{k},t),
\end{equation}
where Eq.~(\ref{R4}) is employed.
On the other hand, the right-hand side of Eq.~(\ref{varphix}) becomes
\begin{eqnarray}
&&
{1\over N_{ph}}\sum_\alpha
\left(
\varphi_{\lambda,\boldsymbol{k}\alpha}(t)E_\alpha(\boldsymbol{k})\varphi^*_{\mu\alpha}(t)
+i{d\varphi_{\lambda,\boldsymbol{k}\alpha}(t)\over dt}\varphi^*_{\mu,\boldsymbol{k}\alpha}(t)
\right)
\nonumber\\
&=&
{1\over N_{ph}}\sum_{\alpha,n,n^\prime}
e^{i(n-n^\prime)\omega t}
(E_\alpha(\boldsymbol{k})-n\omega)
C_{\lambda n,\boldsymbol{k}\alpha}C^*_{\mu n^\prime,\boldsymbol{k}\alpha}
\nonumber\\
&=&
{1\over N_{ph}}\sum_{\alpha,n,N}
e^{-iN\omega t}
E_{\alpha(-n)}(\boldsymbol{k})
C_{\lambda 0,\boldsymbol{k}\alpha(-n)}C^*_{\mu N,\boldsymbol{k}\alpha(-n)}
\nonumber\\
&=&
{1\over N_{ph}}\sum_{\alpha,n}
E_{\alpha(-n)}(\boldsymbol{k})
C_{\lambda 0,\boldsymbol{k}\alpha(-n)}\varphi^*_{\mu,\boldsymbol{k} \alpha(-n)}(t)
\nonumber\\
&=&
{1\over N_{ph}}\sum_{n}
\sum_\alpha
E_{\alpha}(\boldsymbol{k})
C_{\lambda 0,\boldsymbol{k}\alpha}\varphi^*_{\mu,\boldsymbol{k} \alpha}(t)
\nonumber\\
&=&
\sum_\alpha
E_{\alpha}(\boldsymbol{k})
C_{\lambda 0,\boldsymbol{k}\alpha}\varphi^*_{\mu,\boldsymbol{k} \alpha}(t),
\end{eqnarray}
where Eqs.~(\ref{Avarphi2}) and (\ref{AC1}) are employed.
Thus, one obtains 
\begin{equation}
H_{F\mu\lambda}(\boldsymbol{k},t)=\sum_\alpha
E_{\alpha}(\boldsymbol{k})
C_{\lambda 0,\boldsymbol{k}\alpha}\varphi^*_{\mu,\boldsymbol{k} \alpha}(t),
\label{Hx}
\end{equation}
and this is also rewritten as
\begin{eqnarray}
H_{F \lambda\mu}(\boldsymbol{k},t)&=&H^*_{F\mu\lambda}(\boldsymbol{k},t)
\nonumber\\
&=&\sum_\alpha
E_{\alpha}(\boldsymbol{k})
C^*_{\mu 0,\boldsymbol{k}\alpha}\varphi_{\lambda,\boldsymbol{k} \alpha}(t).
\label{Hx2}
\end{eqnarray}

Finally, the following relation is noted as:
\begin{equation}
I_\mu(-1)^nC_{\mu n,\boldsymbol{k}\alpha}=C_{\mu n,-\boldsymbol{k}\alpha}\tilde{I}_{\boldsymbol{k}\alpha},
\label{Cnk}
\end{equation}
where $I_\mu$ and $\tilde{I}_{\boldsymbol{k}\alpha}$ represent the parity of band $\mu$, and a constant phase factor with $|\tilde{I}_{\boldsymbol{k}\alpha}|=1$, respectively.
For more detail, consult Supplemental Materials.\cite{SM2}.
In particular, $\tilde{I}_{\boldsymbol{k}_0\alpha}$ plays a role of the parity of the time-glide spatial inversion\cite{Zhang2021,Morimoto2017} at $\boldsymbol{k}=\boldsymbol{k}_0$ where $\boldsymbol{k}_0$ represents a high-symmetry momentum equal to $\boldsymbol{G}/2$ with $\boldsymbol{G}$ a reciprocal lattice vector satisfying $\boldsymbol{G}\cdot\boldsymbol{R}=2n\pi$ with $n$ an integer.

\section{Derivation of Eq.~(\ref{Phia})}
\label{app2}

Equation~(\ref{H3}) is rewritten in the Bloch representation, where
the Wannier state $|\mu\boldsymbol{R}\rangle$ is
represented in terms of the Bloch state $|\mu\boldsymbol{k}\rangle$
as
\begin{equation}
|\mu\boldsymbol{R}\rangle ={1\over \sqrt{N}}\sum_{\boldsymbol{k}}e^{-i\boldsymbol{k}\cdot\boldsymbol{R}}|\mu\boldsymbol{k}\rangle,
\label{wf}
\end{equation}
with $\boldsymbol{k}$ as the Bloch momentum and $N$ the number of the sites.
In light of the fact that
Eq.~(\ref{H3}) is expressed as
\begin{eqnarray}
&&\sum_{\mu_1\boldsymbol{R}_1 \mu_2\boldsymbol{R}_2 \mu^\prime\boldsymbol{k}^\prime}
\langle \mu\boldsymbol{k}|\mu_1\boldsymbol{R}_1\rangle 
e^{-i\boldsymbol{A}(t)\cdot\boldsymbol{R}_1}
\tilde{h}_{\mu_1\boldsymbol{R}_1,\mu_2\boldsymbol{R}_2}(t)
e^{i\boldsymbol{A}(t)\cdot\boldsymbol{R}_2}
\nonumber\\
&&
\times
\langle \mu_2\boldsymbol{R}_2|\mu^\prime\boldsymbol{k}^\prime
\rangle \langle \mu^\prime\boldsymbol{k}^\prime|\hat{\Phi}\rangle 
=i{\partial\over \partial t}
\langle \mu\boldsymbol{k}|\hat{\Phi}\rangle,
\end{eqnarray}
one obtains the following equation as
\begin{equation}
\sum_{\mu^\prime\boldsymbol{k}^\prime}
\langle \mu\boldsymbol{K}(t)|\tilde{h}(t)|\mu^\prime\boldsymbol{K^\prime}(t)
\rangle \hat{\Phi}_{\mu^\prime\boldsymbol{k}^\prime}
=i{\partial\over \partial t}\hat{\Phi}_{\mu\boldsymbol{k}},
\label{H4}
\end{equation}
with
\(\displaystyle
\boldsymbol{K}(t)\equiv\boldsymbol{k}+\boldsymbol{A}(t), 
\boldsymbol{K}^\prime(t)\equiv\boldsymbol{k}^\prime+\boldsymbol{A}(t)
\)
and 
\(\displaystyle
\hat{\Phi}_{\mu\boldsymbol{k}}\equiv \langle \mu\boldsymbol{k}|\hat{\Phi}\rangle,
\)
where in view of Eq.~(\ref{wf}),
the relations of
\begin{equation}
\sum_{\mu_1\boldsymbol{R}_1}
\langle \mu\boldsymbol{k}|\mu_1\boldsymbol{R}_1\rangle \langle \mu_1\boldsymbol{R}_1|
e^{-i\boldsymbol{A}(t)\cdot\boldsymbol{R}_1}
=\langle \mu\boldsymbol{K}(t)|
\end{equation}
and 
\begin{equation}
\sum_{\mu_2\boldsymbol{R}_2}
e^{i\boldsymbol{A}(t)\cdot\boldsymbol{R}_2}|\mu_2\boldsymbol{R}_2\rangle \langle \mu_2\boldsymbol{R}_2|\mu^\prime\boldsymbol{k}^\prime
\rangle
=|\mu^\prime\boldsymbol{K}^\prime(t) \rangle 
\end{equation}
are employed.
Furthermore, the Hamiltonian matrix element of Eq.~(\ref{H4}) becomes
\begin{eqnarray}
&&\langle \mu\boldsymbol{K}(t)|\tilde{h}(t)|\mu^\prime\boldsymbol{K^\prime}(t)
\rangle 
\nonumber\\
&=&
\int d\boldsymbol{r}\int d\boldsymbol{r}^\prime e^{-i\boldsymbol{K}\cdot\boldsymbol{r}}U^*_{\mu\boldsymbol{K}(t)}(\boldsymbol{r})
\nonumber\\
&&\hspace{2cm}\times
\tilde{h}(\boldsymbol{r},\boldsymbol{r}^\prime;t)
e^{i\boldsymbol{K}^\prime\cdot\boldsymbol{r}^\prime}U_{\mu^\prime\boldsymbol{K}^\prime(t)}(\boldsymbol{r}^\prime)
\nonumber\\
&\approx&
\langle U_{\mu\boldsymbol{0}}(\boldsymbol{r})|
e^{-i\boldsymbol{K}(t)\cdot\boldsymbol{r}}h^0(\boldsymbol{r})
e^{i\boldsymbol{K}(t)\cdot\boldsymbol{r}}|U_{\mu^\prime\boldsymbol{0}}(\boldsymbol{r})
\rangle
\delta_{\boldsymbol{k}\boldsymbol{k}^\prime}
\nonumber\\
&&
+
\int d\boldsymbol{r}\int d\boldsymbol{r}^\prime
e^{-i\boldsymbol{K}\cdot\boldsymbol{r}}U^*_{\mu\boldsymbol{0}}(\boldsymbol{r})
\nonumber\\
&&\hspace{2cm}\times
\boldsymbol{F}(t)\cdot\boldsymbol{\chi}(\boldsymbol{r},\boldsymbol{r}^\prime)
e^{i\boldsymbol{K}^\prime\cdot\boldsymbol{r}^\prime}U_{\mu^\prime\boldsymbol{0}}(\boldsymbol{r}^\prime)
\nonumber\\
&=&
\langle \chi_{\mu}(\boldsymbol{r})|
e^{-i\boldsymbol{K}(t)\cdot\boldsymbol{r}}h(\boldsymbol{r},t)
e^{i\boldsymbol{K}(t)\cdot\boldsymbol{r}}|\chi_{\mu^\prime}(\boldsymbol{r})
\rangle
\delta_{\boldsymbol{k}\boldsymbol{k}^\prime}.
\label{kp}
\end{eqnarray}
In the first equality, Eq.~(\ref{tildeh}) is employed, and the periodic part of the Bloch wavefunction $\langle\boldsymbol{r}|\mu\boldsymbol{K}(t)\rangle$ is represented as $U_{\mu\boldsymbol{K}(t)}(\boldsymbol{r})$.
In the last two equalities, this is replaced by
\begin{equation}
U_{\mu\boldsymbol{0}}(\boldsymbol{r})={1\over \sqrt{N}}\sum_{\boldsymbol{R}}\chi_\mu(\boldsymbol{r}-\boldsymbol{R}), 
\label{U0}
\end{equation}
following the conventional recipe of the $\boldsymbol{k}\cdot\boldsymbol{p}$ perturbation theory, where a Wannier function is introduced as $\chi_\mu(\boldsymbol{r})=\langle\boldsymbol{r}|\mu\rangle$, and $\langle\boldsymbol{r}|\mu\boldsymbol{R}\rangle$ is represented as $\langle\boldsymbol{r}|\mu\boldsymbol{R}\rangle=\langle\boldsymbol{r}-\boldsymbol{R}|\chi_\mu\rangle=\chi_\mu(\boldsymbol{r}-\boldsymbol{R})$.
The resulting expression is in agreement with $H_{\mu\mu^\prime}(\boldsymbol{k},t)\delta_{\boldsymbol{k}\boldsymbol{k}^\prime}$ of Eq.~(\ref{Eqahat}).
By comparing Eq.~(\ref{H4}) with Eq.~(\ref{Eqahat}), one comes to the conclusion that $\hat{\Phi}_{\mu\boldsymbol{k}}(t)=\hat{a}_{\boldsymbol{k}\mu}(t)$ given in Eq.~(\ref{Phia}).

\section{Expressions for the surface magnetization}
\label{app3}
Equations~(\ref{Mbar}) and (\ref{Mbar2}) are rewritten to conform to the surface magnetization.
First, the operator $\hat{a}_{\boldsymbol{k}\mu}(t)$ of Eq.~(\ref{Phia}) is read as
\begin{eqnarray}
\hat{a}_{\boldsymbol{k}\mu}(t)
&=&
\langle \mu\boldsymbol{k}|\hat{\Phi}(t)\rangle
\nonumber\\
&=&
\sum_{\bar{\boldsymbol{k}}^\prime R_y} \langle \mu\boldsymbol{k}|\mu\bar{\boldsymbol{k}}^\prime R_y\rangle\langle\mu\bar{\boldsymbol{k}}^\prime R_y|\hat{\Phi}(t)\rangle
\nonumber\\
&=&
{1\over \sqrt{N_y}}\sum_{R_y}e^{-ik_yRy}\hat{\alpha}_{\bar{\boldsymbol{k}}R_y\mu}(t)
\end{eqnarray}
with $N_y$ the number of sites in the $y$ direction,
where an alternative operator in the partial site-representation
$\hat{\alpha}_{\bar{\boldsymbol{k}}R_y\mu}(t)$ is defined as $\langle\mu\bar{\boldsymbol{k}}R_y|\hat{\Phi}(t)\rangle$.
Thus, the density matrix of Eq.~(\ref{rhoahat}) is recast into
\begin{eqnarray}
\rho_{\boldsymbol{k}\mu\nu}(t)
&=&
{1\over N_y}\sum_{R_y \Delta R_y}
e^{-ik_y\Delta Ry}
\nonumber\\
&&\times
\langle
\hat{\alpha}^\dagger_{\bar{\boldsymbol{k}}R_y\mu}(t)\hat{\alpha}_{\bar{\boldsymbol{k}} (R_y-\Delta R_y)\nu}(t)
\rangle_g.
\end{eqnarray}

The spatial average density matrix in the bulk is given by
\begin{equation}
\eta^{(b)}_{\mu\nu}(t)
=
{1\over V}\sum_{\boldsymbol{k}}\rho_{\boldsymbol{k}\mu\nu}(t)
=
{1\over V}\sum_{\bar{\boldsymbol{k}}R_y}
\rho^\prime_{\bar{\boldsymbol{k}}R_y\mu\nu}(t),
\label{etabC}
\end{equation}
where 
\begin{equation}
\rho^\prime_{\bar{\boldsymbol{k}}R_y\mu\nu}(t)
=
\langle
\hat{\alpha}^\dagger_{\bar{\boldsymbol{k}}R_y\mu}(t)\hat{\alpha}_{\bar{\boldsymbol{k}} R_y\nu}(t)
\rangle_g,
\label{rhobarkRy}
\end{equation}
and the summation over $\boldsymbol{k}$ is taken in the whole BZ in the first equality.
On the other hand, it is necessary to evaluate the corresponding density matrix in the surface, defined as
\begin{equation}
\eta^{(s)}_{\mu\nu}(t)
={1\over S_{xz}}\sum_{\bar{\boldsymbol{k}}}
\Big(\rho^\prime_{\bar{\boldsymbol{k}}0\mu\nu}(t)+\rho^\prime_{\bar{\boldsymbol{k}}L_y\mu\nu}(t)
\Big)
\label{etasC}
\end{equation}
with $S_{xz}$ a surface area in the $x-z$ plane, where the summation over $\bar{\boldsymbol{k}}$ is taken in the limited area of the two-dimensional BZ where surface states are present.
Just both outermost sites in the $y$ direction, namely, $R_y=0$ and $L_y$, are taken in the density matrix of Eq.~(\ref{rhobarkRy}); $L_y$ represents the size of the crystal in the $y$ direction.
It is remarked that $\rho^\prime_{\bar{\boldsymbol{k}}R_y\mu\nu}(t)$ is given by converting Eqs.~(\ref{rhoahat2})-(\ref{OBEoffD}) to the corresponding expressions in the partial-site representation.

Eventually, the surface magnetization is provided by 
\begin{equation}
\bar{\boldsymbol{M}}^{(sol)}(t)
\approx
\mu_B\Re
\sum_{\mu\nu}
\eta^{(s)}_{\mu\nu}(t)
\langle\mu|g_e\boldsymbol{s}+\boldsymbol{l}+\boldsymbol{r}
\times\boldsymbol{A}(t)|\nu\rangle.
\label{MsolsurfC}
\end{equation}
Hence, it is found that this arises from the spin magnetization $(g_e\boldsymbol{s})$, the orbital magnetization $(\boldsymbol{l})$ relevant to the IFE, and the remaining term $(\boldsymbol{r}
\times\boldsymbol{A}(t))$ due to the gauge correction.
Here, it is noted that the term 
$
\sum_{\boldsymbol{k}}
\rho_{\boldsymbol{k}\mu\nu}(t)
\langle\mu|\boldsymbol{r}|\nu\rangle
\times\boldsymbol{k},
$
which is seen 
in Eq.~(\ref{Mbar2}), is neglected.
This is because the leading contribution of the polarization coherence $f(t)\equiv \rho_{\boldsymbol{k}\mu\nu}(t)$ with $\mu\not=\nu$ arises from the linear optical process with respect to the laser field $\boldsymbol{F}(t)$, and most of the dipole transitions are effective only when the almost on-resonant condition of 
$\omega \approx E_g \approx \Delta_{\mu\nu}(\boldsymbol{k})$ is met, namely, in the vicinity of the $\Gamma$ point, where $\Delta_{\mu\nu}(\boldsymbol{k})$ represents the energy difference between bands $\mu$ and $\nu$.
This statement leads to the assumption that the associated induced magnetization vanishes in the asymptotic limit of time.

Below, this is verified in more detail.
The procedure of taking the infinite limit of $t$ on the non-singular function $f(t)$ in the interval $[0,\infty)$ is replaced by another procedure of taking a limit of an infinitesimal value of $\epsilon$, which is given by
\begin{equation}
\lim_{t\rightarrow\infty} f(t)
=\lim_{\epsilon\rightarrow +0}
\epsilon\int^\infty_0 dt\: e^{-\epsilon t}f(t).
\end{equation}
This equality is simply verified by taking a partial differentiation with respect to $t$ in the right-hand side.
Given 
\begin{equation}
f(t)=\sum_n\:e^{i\omega_n t}f_n
\end{equation}
with $\omega_n\not=0$ for $n\not=0$ and $\omega_0=0$, it is shown that
\begin{equation}
\lim_{t\rightarrow\infty} f(t)
=\lim_{\epsilon\rightarrow +0}\sum_n f_n\:
{\epsilon\over 
\epsilon-i\omega_n}
=f_0,
\label{limtinf}
\end{equation}
implying that all of the
oscillating terms of $f(t)$ vanish asymptotically and just a constant contribution $f_0$ is extracted.
Since the leading term of $f(t)$ is linear in $\boldsymbol{F}(t)$, as stated above, $f_0$ is considered to almost vanish.
Besides, it is remarked that there would possibly remain small nonzero contributions due to higher-order optical processes just in the vicinity of $\boldsymbol{k}=\boldsymbol{0}$.
Therefore, it is concluded that 
$
\sum_{\boldsymbol{k}}
\rho_{\boldsymbol{k}\mu\nu}(t)
\langle\mu|\boldsymbol{r}|\nu\rangle
\times\boldsymbol{k}
\approx \boldsymbol{0}
$
in Eq.~(\ref{Mbar2}).


\begin{thebibliography}{00}
\bibitem{comment1}Due to a heavy computational burden, the numerical calculations of the induced bulk magnetization are implemented just in the limited $k_x-k_z$ region, namely, $|k_xd_x|, |k_zd_z| \le 0.1\pi$, in the bulk crystal that is confined in the $y$ direction. The valence electrons with down spin are dominantly excited in the vicinity of the $\Gamma$ point corresponding to the above region. The obtained result is about 0.4 mT at $t=25T$, which is already almost constant in time. Thus, by simply extending the incorporated region of the $k_x-k_z$ plane from the above limited region to the whole region, the net bulk magnetization is estimated to be approximately $(\pi/0.1\pi)^2 \times 0.4$ mT $=40$ mT, though this is merely an ideal maximum. This value is mostly ascribable to the IFE, not to both the spin magnetization and the gauge correction that are negligibly small.
\end{thebibliography}

\end{document}